\documentclass[iop, apj, twocolappendix, numberedappendix, appendixfloats]{emulateapj}
\usepackage{threeparttable}
\usepackage{multirow}
\usepackage{times}
\usepackage{enumitem}
\usepackage{url} 
\usepackage{color} 
\usepackage{amsmath,bm}

\shorttitle{}
\shortauthors{Y. Huang et al.}

\begin{document}

\title{Milky Way tomography with the SkyMapper Southern Survey:\\ I: Atmospheric parameters and distances of one million red giants}

\author{Y. Huang\altaffilmark{1,9}}
\author{ B.-Q. Chen\altaffilmark{1}}
\author{ H.-B. Yuan\altaffilmark{2}}
\author{ H.-W. Zhang\altaffilmark{4}}
\author{ M.-S. Xiang\altaffilmark{3}}
\author{ C. Wang\altaffilmark{4,8}}
\author{H.-F. Wang\altaffilmark{1,7,8}}
\author{C. Wolf\altaffilmark{5}}
\author{G.-C. Liu\altaffilmark{6}}
\author{ X.-W. Liu\altaffilmark{1,9}}

\altaffiltext{1}{South-Western Institute for Astronomy Research, Yunnan University, Kunming 650500, People's Republic of China; {\it yanghuang@ynu.edu.cn {\rm (YH)}; x.liu@ynu.edu.cn {\rm (XWL)}}}
\altaffiltext{2}{Department of Astronomy, Beijing Normal University, Beijing 100875, People's Republic of China}
\altaffiltext{3}{Max-Planck Institute for Astronomy, K{\"o}nigstuhl, D-69117, Heidelberg, Germany}
\altaffiltext{4}{Department of Astronomy, Peking University, Beijing 100871, People's Republic of China}
\altaffiltext{5}{Research School of Astronomy and Astrophysics, Australian National University, Canberra, ACT 2611, Australia}
\altaffiltext{6}{China Three Gorges University, Yichang 443002, People's Republic of China}
\altaffiltext{7}{Department of Astronomy, China West Normal University, Nanchong 637009, China}
\altaffiltext{8}{LAMOST Fellow}
\altaffiltext{9}{Corresponding authors}

\begin{abstract}
Accurate determinations of atmospheric parameters (effective temperature $T_{\rm eff}$, surface gravity log\,$g$ and metallicity [Fe/H]) and distances for large complete samples are of vital importance for various Galactic studies. 
We have developed a photometric method to select red giant stars and estimate their atmospheric parameters from the photometric colors provided by the SkyMapper Southern Survey (SMSS) data release (DR) 1.1,  using stars in common with the LAMOST Galactic spectroscopic surveys as a training set.
Distances are estimated with two different approaches: one based on the Gaia DR2 parallaxes for nearby ($d \leq 4.5$\,kpc) bright stars and another based on the absolute magnitudes predicted by intrinsic color $(g-i)_0$ and photometric metallicity [Fe/H] for distant ($d > 4.5$\,kpc) faint stars.
Various tests show that our method is capable of delivering atmospheric parameters with a precision of $\sim$80\,K for $T_{\rm eff}$, $\sim$0.18\,dex for [Fe/H] and $\sim$0.35\,dex for log\,$g$ but with a significant systematic error at log\,$g \sim$\,2.3.
For distances delivered from $(g-i)_0$ and photometric [Fe/H], our test with the member stars of globular clusters show a median uncertainty of 16 per cent with a negligible zero-point offset.
Using this method, atmospheric parameters and distances of nearly one million red giant stars are derived from SMSS DR1.1.
Proper motion measurements from Gaia DR2 are available for almost all of the red giant stars, and radial velocity measurements from several large spectroscopic surveys are available for 44 per cent of these.
This sample will be accessible  online at \url{https://yanghuang0.wixsite.com/yangh/research}.
\end{abstract}

\section{Introduction}
Understanding the formation and evolution of galaxies is one of the most challenging problems in astrophysics.
As our own galaxy, the Milky Way (MW) provides a unique opportunity to study a galaxy in exquisite detail, by measuring and analyzing the properties (e.g., atmospheric parameters, radial velocities, distances and proper motions) of large samples of constituent stars.

Over the last decades many large Galactic surveys were conducted and revolutionized our knowledge about the MW.
These include photometric ones like the Sloan Digital Sky Survey (SDSS; York et al. 2000), the Two Micron All Sky Survey (2MASS; Skrutskie et al. 2006), the Pan-STARRS1 surveys (PS1; Chambers et al. 2016) and the SkyMapper Southern Survey (SMSS; Wolf et al. 2018), as well as spectroscopic surveys such as RAVE (Steinmetz et al. 2006), SDSS/SEGUE (Yanny et al. 2009), SDSS/APOGEE (Majewski et al. 2017), LAMOST (Deng et al. 2012; Liu et al. 2014) and GALAH (De Silva et al. 2015), and finally astrometric ones such as Hipparcos (Perryman et al. 1997) and Gaia (Gaia Collaboration et al. 2016, 2018a).
However, two important issues limit our further understanding of the MW.
First, while there are hundreds of billions of Galactic stars in the whole sky of $4\pi$ steradians, only a very limited fraction of them can be targeted spectroscopically, which, although rapidly increasing, is currently of the order of about ten million (largely contributed by the LAMOST Galactic surveys).
More importantly, the selection effects of spectroscopic surveys are difficult to evaluate, but needed for recovering the underlying population properly. 
Another issue is that spectroscopic surveys in the southern sky are much smaller and shallower than their counterparts in the North.
For these reasons it is hard to draw a comprehensive picture of and gain unbiased insight into the metallicity and velocity distributions of our MW, which are key to understanding the formation and evolution of the MW as a galaxy. 

The ongoing SMSS can significantly help to address the above issues.
First, it utilizes a set of $uvgriz$ filters designed by Bessell et al. (2011) that is sensitive to stellar atmospheric parameters.
The SkyMapper $u$ band is similar to the Str{\"o}mgren $u$ band but narrower than the SDSS $u$ band, and thus provides photometric sensitivity to stellar surface gravity.
The SkyMapper $v$ band is similar to the DDO\,38 band (McClure \& van den Bergh 1968) and is very sensitive to stellar metallicity, particularly at low metallicities.
With those specially designed filters, the SMSS photometry alone can provide relatively precise measurements of stellar atmospheric parameters of up to a billion stars down to $r \ga 20$\,mag once the survey reaches its final depth for the whole southern sky (Keller et al. 2007; Casagrande et al. 2018). 
SMSS should thus provide a dramatic increase in the number of stars with accurately measured atmospheric parameters (especially metallicity) in the southern sky.  

In this paper, we attempt to single out a clean sample of red giant stars and derive their atmospheric parameters from the recently released SMSS DR1.1 (Wolf et al. 2018, hereafter W18).
Red giant stars are selected in order to have a large distance coverage, given the fact that the current SMSS DR1.1 only contains photometric data from the {\it Shallow Survey}, where $10\sigma$-limiting magnitudes are $\sim 18$ in all bands.
In addition to the atmospheric parameters, distances of the sample stars are also derived, using parallaxes provided by Gaia DR2 for relatively nearby stars (Gaia Collaboration et al. 2018a; Lindegren et al. 2018) and from photometric parallaxes estimated with a likelihood method for more distant stars.
The paper is structured as follows.
In Section\,2, we describe the data used in this paper.
The giant star selection is introduced in Section\,3.
In Section\,4, we develop a method to derive atmospheric parameters and distances for red giant stars based on SMSS photometry.
Various tests of the estimated atmospheric parameters and distances are presented in Section\,5.
We construct the SMSS giant star sample in Section\,6 by applying the selection algorithm and the atmospheric parameter and distance estimation method to the whole SMSS DR1.1.
We briefly introduce the potential applications and perspectives of the sample in Section\,7.
Finally, a brief summary is presented in Section\,8.

Before starting, it is worth mentioning that we have noted an independent work by Casagrande et al. (2019, hereafter C19).
Although we have lots of common purposes, the methods and results between their and our work have significant differences (see Section\,7 for details).
Also, we stress that we are two independent work.

\section{Data}
\begin{figure*}
\begin{center}
\includegraphics[scale=0.45,angle=0]{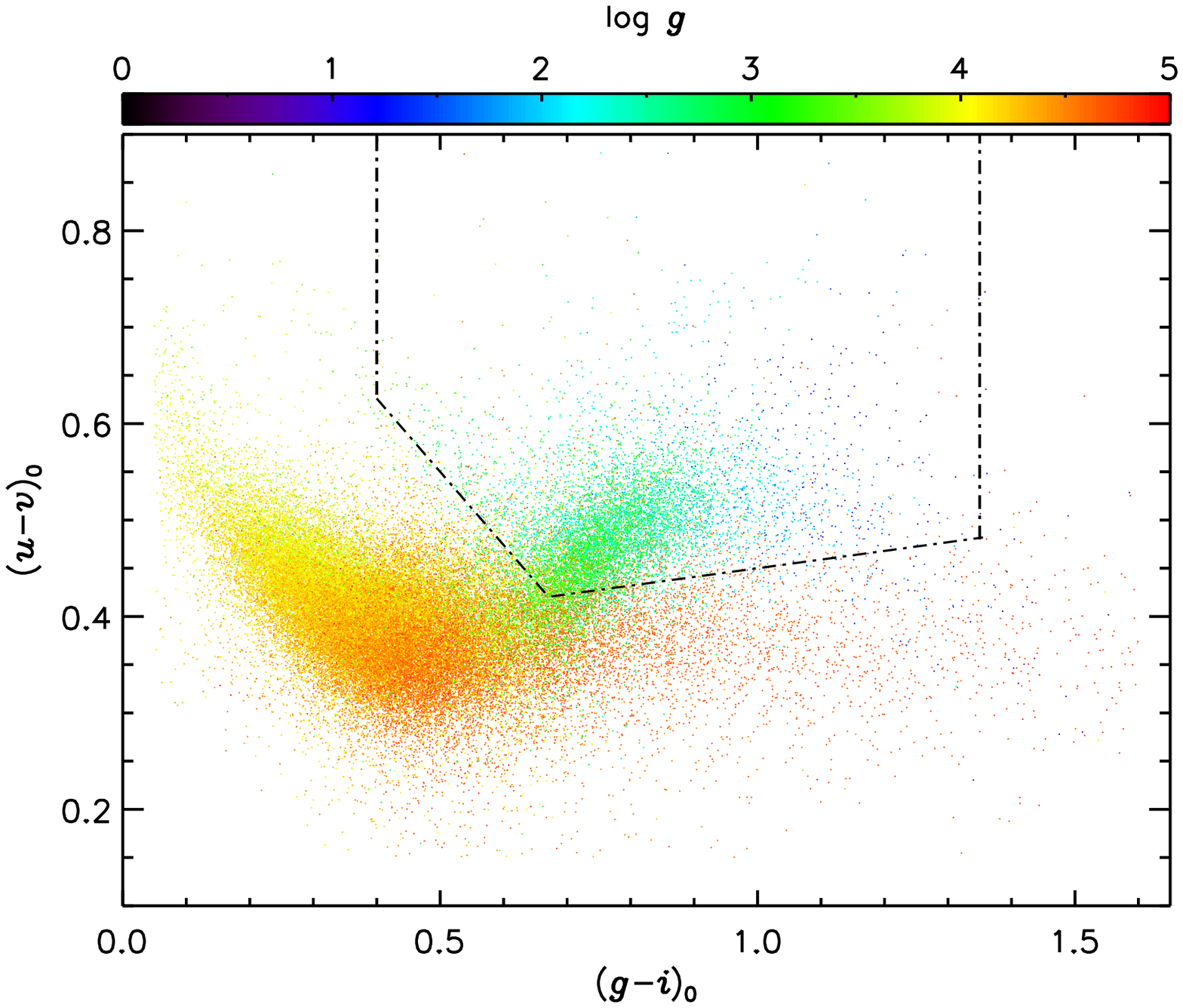}
\includegraphics[scale=0.45,angle=0]{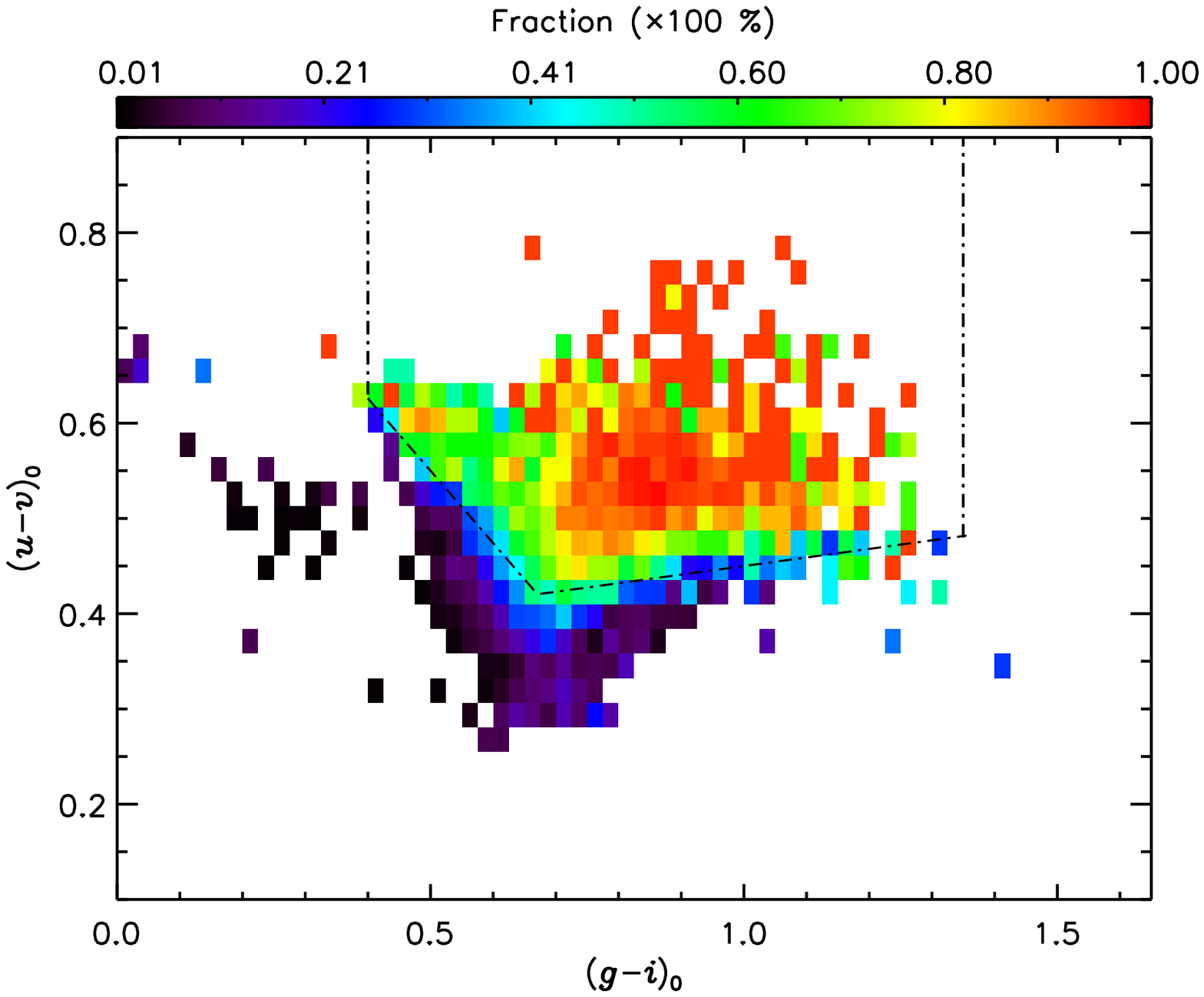}
\caption{{\it Left panel}: $(u-v)_{0}$ versus $(g-i)_{0}$ diagram of the {\it gLS} sample stars, color-coded by the LAMOST surface gravity. {\it Right panel}: Giant fraction distributions in the $(u-v)_{0}$ versus $(g-i)_{0}$ plane (with a binsize of 0.025\,mag in each axis). Dot-dashed lines in both panels denote the cuts, defined by Eqs.\,(1)--(3), that we develop to select the red giant stars. }
\end{center}
\end{figure*}

\begin{figure}
\begin{center}
\includegraphics[scale=0.45,angle=0]{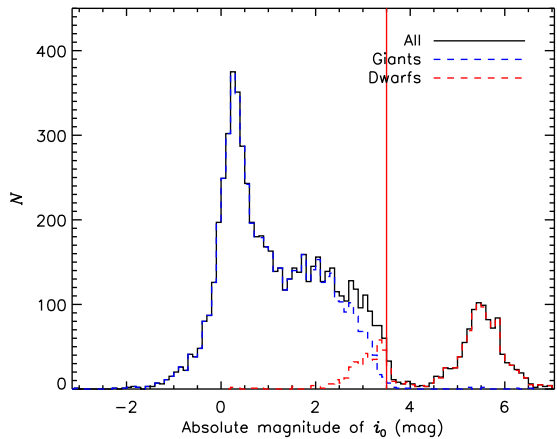}
\caption{Histogram of $M_{i_{0}}$ for 7391 stars selected from the {\it gLS} sample using the color cuts defined by Eqs.\,(1)--(3) combined with the parallax quality ($\sigma_{\varpi}$/$\varpi \le 0.2$) and distance cuts (smaller than 4.5\,kpc). 
The black solid, blue dashed and red dashed lines represent all stars, red giants and dwarfs, respectively.
Note that a few stars with absolute magnitudes smaller than 3.5 are classified as dwarfs due to large uncertainties in their surface gravity estimates.}
\end{center}
\end{figure}
 
In the current work, we use data from the SMSS DR1.1 (W18).
The SMSS is an optical multi-band ($uvgriz$), wide-field survey, aiming to cover the entire Southern hemisphere down to a limiting magnitude of $\sim 22$\,mag in $r$ band.
It uses a 1.35-m telescope located at the Siding Spring Observatory, which is equipped with a 5.7 sq. deg. field-of-view Cassegrain imager and a detector mosaic composed of 32 2k$\times$4k CCDs.
The SMSS was started on March 15 2014, and 
the first year as well as the bright time after Year 1 are used for the {\it Shallow Survey} of short exposures (less than 5\,min), aiming to provide a robust calibration reference and sensitivity to variability.
Since Year 2 most of the observing time is dedicated to the {\it Main Survey}, which plans to survey the Southern hemisphere to 22\,mag by the year 2021.
In December 2017 the SMSS DR1.1 was released for world-wide access, based on over 66\,000 images from the {\it Shallow Survey} that cover 17\,200 sq. deg. of sky and catalog 285 million unique astrophysical objects down to a limiting magnitude of $\sim 18$\,mag in all bands.

In addition to the SMSS DR1.1, we use data from the one-year Pilot surveys (2011 September to 2012 June) and the four-years Regular surveys (2012 September to 2016 June) of the LAMOST Galactic spectroscopic surveys (Xiang et al. 2017b).
Atmospheric parameters are derived with the LSP3 pipeline developed by Xiang et al. (2015, 2017a), and those parameters are now all released and available on \url{http://dr4.lamost.org/doc/vac}.
The data are used to define photometric selection criteria of red giant stars and to calibrate the relation between stellar metallicity and the photometric colors from SMSS DR1.1.
Other spectroscopic surveys, including SDSS/APOGEE DR14 (Abolfathi et al. 2017) and GALAH DR2 (Buder et al. 2018), are also used for parameter validation purposes.
Here, we adopt stellar atmospheric parameters yielded by the low-resolution LAMOST Spectroscopic Survey rather than by the medium/high resolution surveys (e.g. the APOGEE or GALAH surveys) for the following reasons: 
1) The number of stars in common between LSS-GAC DR3 and SMSS DR1.1 runs into a few hundred thousand (see Section\,3), which is much more numerous than samples from other spectroscopic surveys (APOGEE, GALAH); 
2) The LAMOST surveys ($r \sim$\,$17$--$18$\,mag) are much deeper than the medium/high resolution surveys ($H =13.8$\,mag for the APOGEE survey and $V = 14$\,mag for the GALAH survey), providing more halo stars of low metallicities in our calibration sample (see Section\,3) for training and calibration;
3) The typical uncertainties of metallicity and surface gravity of red giant stars when measured with the stellar parameter determination pipeline LSP3 in LAMOST spectra with signal-to-noise ratios (SNRs) better than 30 using, are better than 0.1\,dex (Xiang et al. 2017b), comparable to results yielded by medium/high resolution spectroscopic surveys.

Finally, we also use parallaxes, proper motions and radial velocities released in Gaia DR2 (Gaia Collaboration et al. 2018), and radial velocities from SDSS/SEGUE DR12 (Alam et al. 2012) and RAVE DR5 (Kunder et al. 2017).
Unless specified otherwise, all magnitudes and colors presented here refer to dereddened values, corrected using reddening values taken from the extinction map\footnote{We have corrected a $\sim$15.5 per cent systematic overestimated by Schlegel, Finkbeiner \& Davis (1998), as found by previous work (e.g. Schlafly et al. 2010; Berry et al. 2012).} of Schlegel, Finkbeiner \& Davis (1998; hereafter SFD98) and extinction coefficients derived here for SkyMapper bands (see Appendix\,A for details) and using those of Yuan, Liu \& Xiang (2013) for other bands.

\section{Giant star selection}
In this Section, we develop photometric criteria to select red giant stars using SkyMapper colors and Gaia parallaxes, guided by values of surface gravity measured from the LAMOST spectra.
For this purpose, we first cross-match SMSS DR1.1 with LSS-GAC DR3 and find 257,866 common stars (hereafter the {\it LS sample}).
For calibration we need atmospheric parameters and photometric colors of sufficiently high precision and thus apply the following cuts to the {\it LS} sample:
\begin{itemize}[leftmargin=*]

\item The stars must have a Galactic latitude $|b| \ge 30^{\circ}$, in order to minimise uncertainties due to reddening corrections;

\item The stars must have LAMOST spectra with SNR greater than 10 and effective temperatures below 10000\,K;

\item The photometric errors in $uvgi$ bands must smaller than 0.035\,mag.

\end{itemize}
After these cuts, 72\,384 stars remain in what we now call the golden {\it LS} sample (hereafter {\it gLS sample}).
We show in the left panel of Fig.\,1 the distribution of {\it gLS} stars in the $(g-i)_0$ versus $(u-v)_0$ color-color diagram, color-coded by surface gravity log\,$g$.
Here, the color $(g-i)_0$ is an indicator of effective temperature $T_{\rm eff}$ and the color $(u-v)_0$ is believed to be sensitive to log\,$g$ (Bessell et al. 2011).
As the plot shows, red giant stars (defined by log\,$g \le 3.5$ hereafter) are mostly found in the upper parts of the color-color diagram and clearly separated from cool dwarf stars (defined by log\,$g > 3.5$ hereafter).
We find that 14 per cent (9982/72\,384) of the stars in the gLS sample are giants.
To quantitatively define the color selection criteria of red giant stars, we further show the distribution of giant fraction in the right panel of Fig.\,1. 
We then empirically define the color cuts that follows roughly the 60 per cent contour of the number ratios. 
This yields,

\begin{equation}
(u-v)_{0} \ge 0.93 - 0.76\times(g-i)_{0},
\end{equation}
\begin{equation}
(u-v)_{0} \ge 0.36 + 0.09\times(g-i)_{0},
\end{equation}
\begin{equation}
0.40 \le (g-i)_{0} \le 1.35.
\end{equation}
With the above cuts, 72 per cent (7140/9982) of all giants in the gLS sample are successfully selected, along with a 17 per cent (1495/8635) contamination from dwarf stars.

\begin{table}
\centering
\caption{Fit Coefficients}
\begin{threeparttable}
\begin{tabular}{lcc}
\hline
Coeff. & [Fe/H]\tnote{a} & log\,{\it g} \tnote{b}\\
\hline
$a_{0}$&$-1.80830$&$+7.14971$\\
$a_{1}$&$+3.59975$&$+4.2838$\\
$a_{2}$&$-5.03963$&$-12.83067$\\
$a_{3}$&$+3.63297$&$-6.38869$\\
$a_{4}$&$-1.03745$&$-3.59100$\\
$a_{5}$&$-4.63564$&$-7.39646$\\
$a_{6}$&&$+17.22680$\\
$a_{7}$&&$+1.11532$\\
$a_{8}$&&$+2.27537$\\
$a_{9}$&&$+2.85317$\\
\hline
\end{tabular}
\begin{tablenotes}
\item[a] The fitting function is given by Eq.\,(4).
\item[b] The fitting function is given by Eq.\,(5).
\end{tablenotes}
\end{threeparttable}
\end{table}

Thanks to the accurate parallax measurements provided by Gaia DR2, most of those dwarf contaminators could be removed by applying an absolute magnitude cut.
In Fig.\,2 we show the resulting histogram of values of absolute magnitude in $i$ band, $M_{i_{0}}$, for 7391 stars that pass the above color cuts and have good-quality parallax measurements ($\sigma_{\varpi}$/$\varpi \le 0.2$). 
Here, rather than estimating distances by simply inverting the Gaia parallax measurements, we adopt the distance estimates\footnote{Stars of distances greater than 4.5\,kpc are discarded considering the non-negligible systematics in the distances (Huang et al. in preparation) estimated by Bailer-Jones et al. (2018).} of Bailer-Jones et al. (2018), who provide distances of 1.33 billion stars derived from the Gaia parallaxes by applying a weak prior on the distribution of Galactic stars. 
We note that stars without good-quality parallax measurements (1244 in number) are likely red giant stars given their likely large distances.
Therefore, as Fig.\,2 shows, a cut of $M_{i_0} \ge 3.5$\,mag should exclude most of the dwarf stars and suppress the contamination to a level of a few per cent.

Finally, with the above color and absolute magnitude cuts, 7629 stars (7125 giants and 504 dwarfs) are left.
In summary, the criteria developed above allow us to single out red giant stars by combining SkyMapper colors and Gaia parallaxes, with a completeness of $\sim 71$ per cent and a purity of $\gtrsim 93$ per cent.

\section{Atmospheric Parameters and distance Estimations}
\begin{figure}
\begin{center}
\includegraphics[scale=0.5,angle=0]{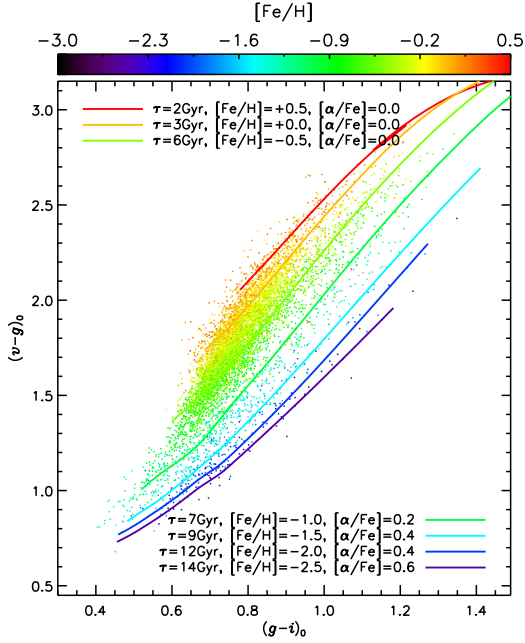}
\caption{$(v - g)_0$ versus $(g - i)_0$ diagram of red giant stars (7140) from the {\it gLS} sample stars, color-coded by LAMOST metallicity.
Lines of different colors are the predicted $(v - g)_0$ versus $(g - i)_0$ sequences of red giant stars (log\,$g \leq 3.5$) for different values of metallicity, age and [$\alpha$/Fe] from the Dartmouth Stellar Evolution Database (Dotter et al. 2008).}
\end{center}
\end{figure}

\begin{figure*}
\begin{center}
\includegraphics[scale=0.48,angle=0]{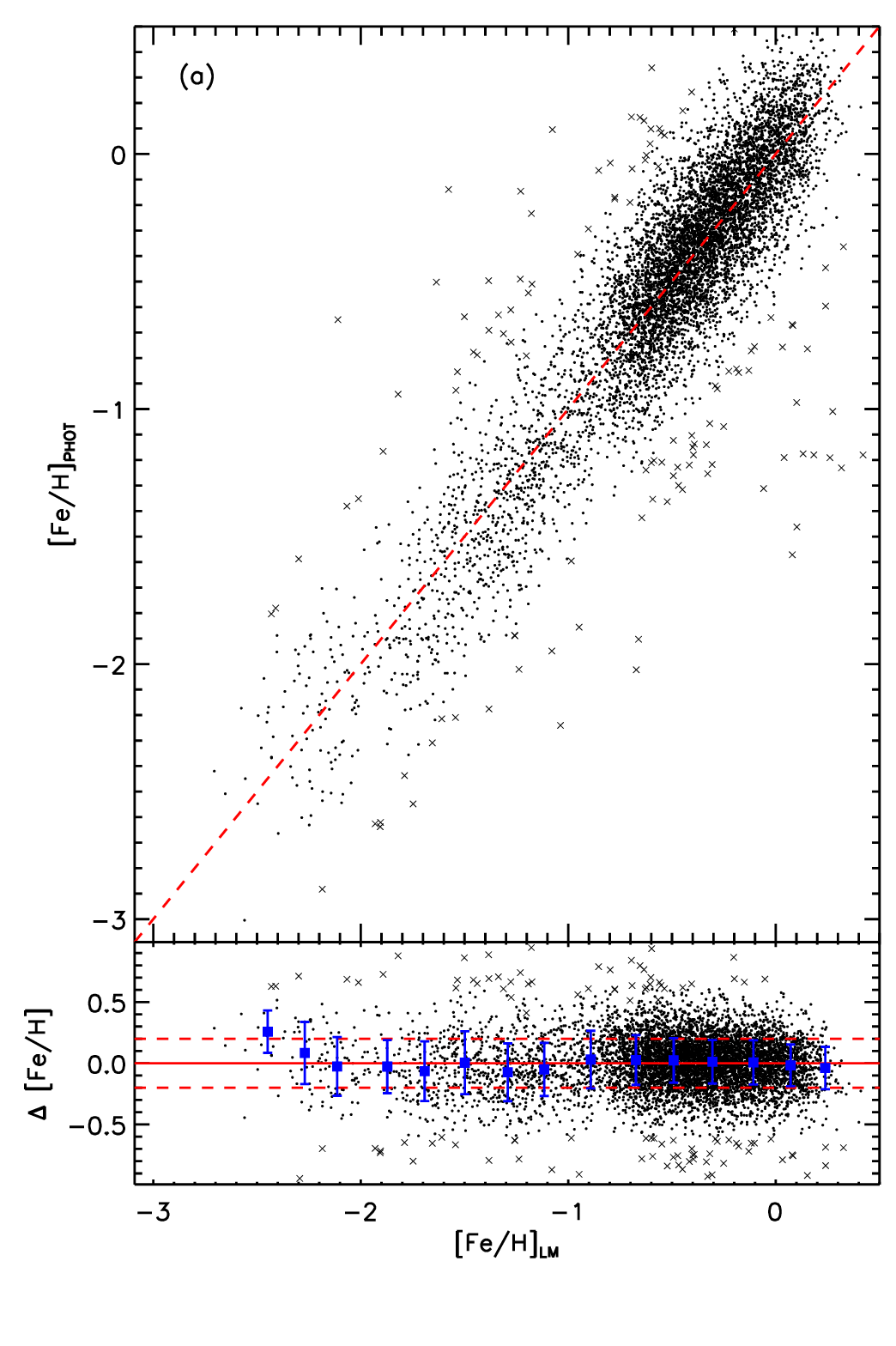}
\includegraphics[scale=0.48,angle=0]{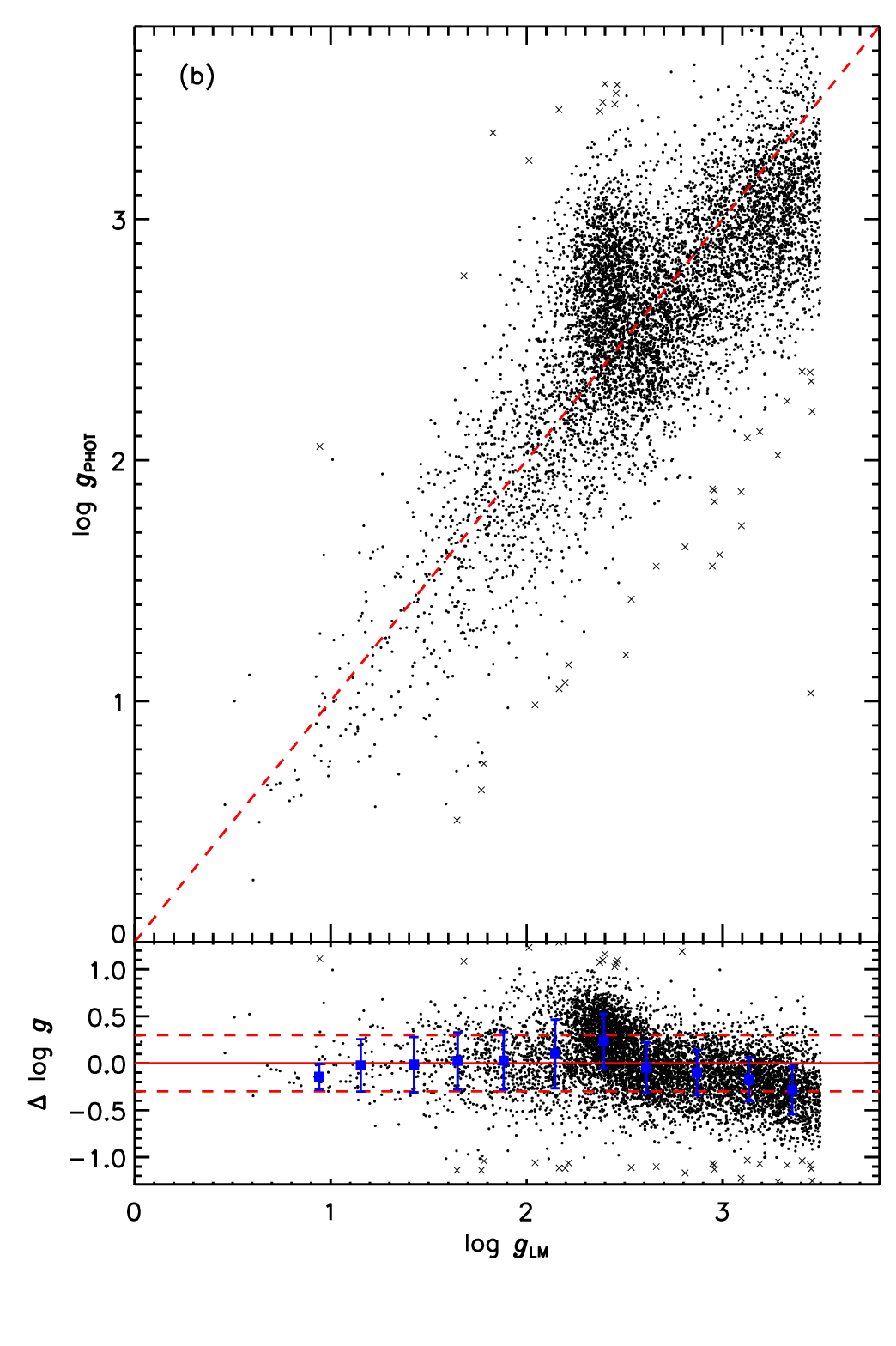}
\caption{{\it Left}: The top panel shows spectroscopic versus photometric metallicities with the latter calculated using the fit presented in Section\,4.1, which is derived using over 7000 red giant stars in the {\it gLS} sample. 
             The bottom shows the residuals ([Fe/H]$_{\rm PHOT} -$\,[Fe/H]$_{\rm LM}$) as a function of spectroscopic metallicity.
             The blue dots and error bars represent the means and standard deviations of the residuals in the individual spectroscopic metallicity bins.
             Red dashed lines are the $1\sigma$ scatter of the residuals and the red solid line marks zero residuals.
             {\it Right}: Same as the left panel but for surface gravity.
             In both panels, crosses are data points excluded by 3$\sigma$ clipping and dots represent those adopted in the fitting.}
\end{center}
\end{figure*}

\begin{table*}2
\centering
\caption{Parameters of the fiducial clusters}
\begin{threeparttable}
\begin{tabular}{ccccccc}
\hline
Name & $l$& $b$& $E(B-V)$ & DM &[Fe/H]&References\tnote{a}\\
&(degree)&(degree)&&&&\\
\hline
NGC\,6791&$69.958$&$+10.904$&$0.10$&$13.26$&$+0.40$&1,2\\
NGC\,6838&$56.744$&$-4.564$&$0.28$&$13.73$&$-0.81$&2-5\\
NGC\,5904&$3.863$&$+46.796$&$0.03$&$14.35$&$-1.26$&2, 3, 5\\
NGC\,6205&$59.008$&$+40.912$&$0.02$&$14.40$&$-1.60$&2, 3, 5\\
NGC\,7099&$27.179$&$-46.836$&$0.03$&$14.66$&$-2.29$&2, 3, 5, 6, 7\\
NGC\,7078&$53.371$&$-35.770$&$0.10$&$15.41$&$-2.42$&2, 3, 5\\
\hline
\end{tabular}
\begin{tablenotes}
\item[a] References: 1 -- WEBDA (https://www.univie.ac.at/webda/); 2 -- An et al. (2009); 3 --Harris (2010); 4 -- Grundahl et al. (2002); 5 -- Kraft \& Ivans (2003); 6 -- Kains et al. (2013); 7 -- O'Malley \& Chaboyer (2018)
\end{tablenotes}
\end{threeparttable}
\end{table*}

In this Section, we present empirical relations between atmospheric parameters and photometric colors for red giant stars.
Also, a likelihood method is developed for estimating absolute magnitudes of red giant stars, using an empirically calibrated color--absolute magnitude fiducial.

\subsection{Metallicity}
To show the sensitivity of SkyMapper color $(v-g)_0$ to stellar metallicity, 7140 red giants from the gLS sample are shown in $(v-g)_0$ versus $(g-i)_0$ diagram in Fig.\,3, color-coded by stellar metallicities measured from LAMOST spectra (Xiang et al. 2015, 2017ab).
The plot clearly shows metallicity sequences of different metallicities ranging from $-2.5$ to $0.5$\,dex as $(v-g)_0$ changes for the full $(g-i)_0$ color range (spanning roughly spectral types from late-G to late-K giants).
$(v-g)_0$ versus $(g-i)_0$ sequences for metallicities (and of different ages and [$\alpha$/Fe] ratios) predicted by the Dartmouth Stellar Evolution Database (Dotter et al. 2008) are shown to be in good agreement with the observations.
The sensitivity of $(v-g)_0$ to [Fe/H] is about 0.3-0.4\,mag per dex, which provides much better sensitivity than the widely used metallicity estimator $(u-g)_0$ from SDSS photometry (e.g. Ivezi{\'c} et al. 2008; Yuan et al. 2015ab).

For estimating photometric metallicities from SkyMapper colors, we carry out a two-dimensional second-order polynomial fit of spectroscopic metallicity [Fe/H] as a function of color $(v-g)_0$ and $(g-i)_0$, using the 7140 red giants selected from the gLS sample,
\begin{equation}
 \begin{split}
{\rm [Fe/H]} =\,&a_{0} + a_{1}X_{1}+ a_{2}X_{2} +\\
                    &a_{3}X_{1}X_{2} + a_{4}X_{1}^{2} + a_{5}X_{2}^{2}\text{,}
\end{split}
\end{equation}
where $X_1$ and $X_2$ denote $(v-g)_0$ and $(g-i)_0$, respectively, and $a_i$ ($i = 0,..., 5$) are fit coefficients.
Three-sigma clipping is performed in the fitting process.
We have checked that inclusion of other color terms or using higher order polynomials do not improve the fit significantly.
The resulting fit coefficients are listed in Table\,1.
A comparison of the input spectroscopic metallicities and the photometric estimates is shown in the left panel of  Fig.\,4.
One sees that the photometric metallicities from the fit in Eq.\,(4) match the input spectroscopic values very well.
Fit residual as a function of spectroscopic metallicity are also shown in Fig.\,4, and
the plot shows no obvious systematics for [Fe/H]\,$\ge -2.4$.
Some mild systematic deviations, at the level of about 0.2\,dex, are noticeable at lower metallicities, $-2.6 <$\,[Fe/H]\,$<-2.4$.
The dispersion of the residuals is around 0.17-0.20\,dex.

\begin{figure*}
\begin{center}
\includegraphics[scale=0.5,angle=0]{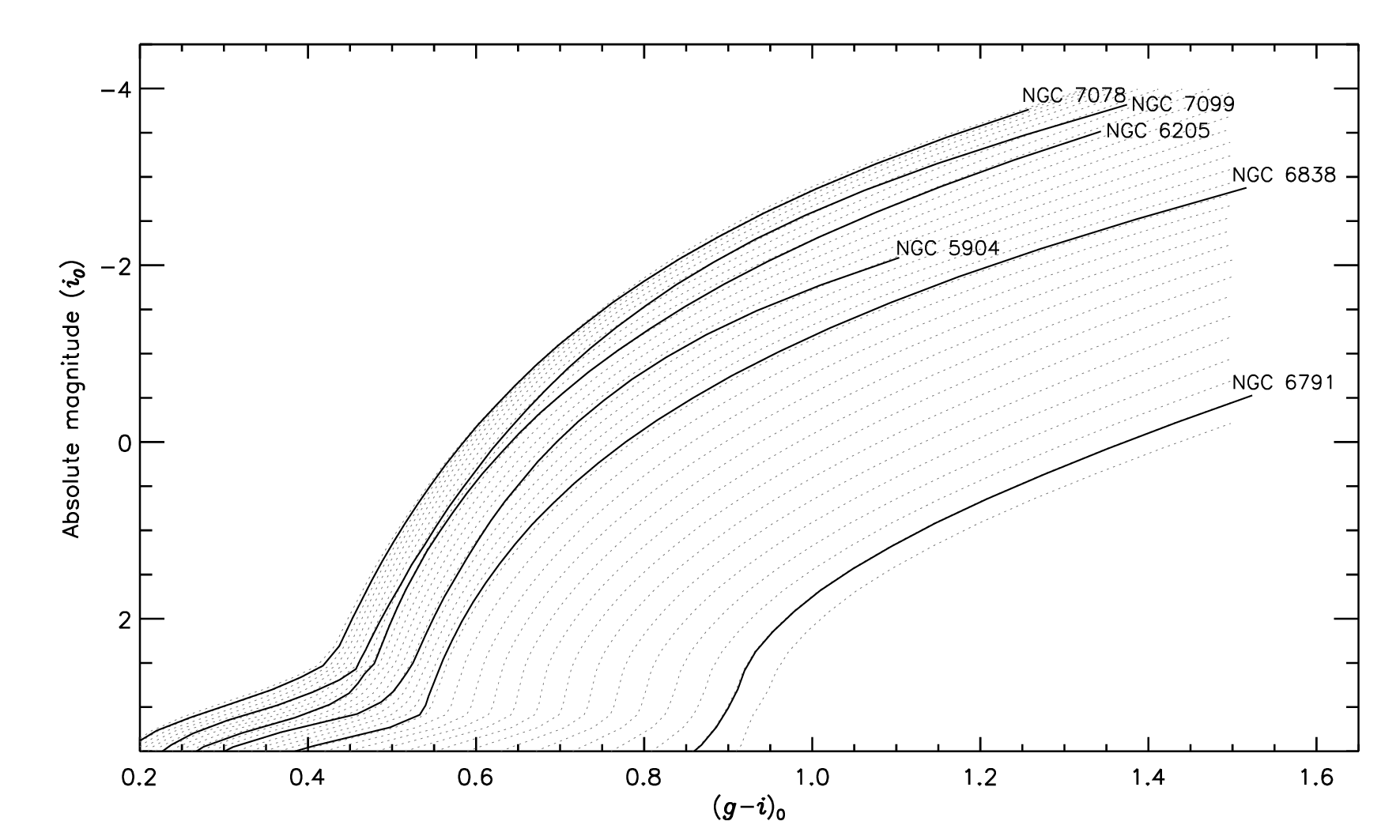}
\caption{Interpolation of the six giant-branch fiducials (thick lines),  $M_{i_{0}}$ versus $(g-i)_0$. The dotted lines show the set of interpolated fiducials.}
\end{center}
\end{figure*}
 
\subsection{Surface gravity}
The sensitivity of  SkyMapper color $(u-v)_0$ to surface gravity is shown in Fig.\,1 and has already been discussed in Section\,3.
Similar to the treatment of metallicity in Section\,4.1, we have carried out a three-dimensional second-order polynomial fit to spectroscopic surface gravity log\,$g$ as a function of colors $(u-v)_0$, $(v-g)_0$ and $(g-i)_0$, using data of the red giants selected from the gLS sample,

\begin{equation}
 \begin{split}
{\rm log}\,g =\,&a_{0} + a_{1}X_{1}+ a_{2}X_{2} + a_{3}X_{3} +\\
                    &a_{4}X_{1}X_{2} + a_{5}X_{1}X_{3} + a_{6}X_{2}X_{3} + \\
                    &a_{7}X_{1}^{2} + a_{8}X_{2}^{2} + a_{9}X_{3}^{2}\text{,}
\end{split}
\end{equation}
where $X_1$, $X_2$ and $X_3$ denote $(v-g)_0$, $(g-i)_0$ and $(u-v)_0$, respectively, and $a_i$ ($i = 0,..., 9$) are fit coefficients.
Three-sigma clipping is again applied in the fitting process.
Again, we have checked that inclusion of other color terms or using higher order polynomials  do not improve the fit significantly.
The resultant fit coefficients are listed in Table\,1.
A comparison of the input spectroscopic surface gravities and our the fitted values is shown in the right panel of Fig.\,4.
As the plot shows, the fit is quite reasonable except for giants of log\,$g$ greater than 2.3 (see the bottom panel of the plot).
In particular, the surface gravities of red clump stars (hereafter RCs) with log\,$g \sim 2.3$ are poorly determined by SkyMapper colors as they are systematically overestimated by $\sim 0.25$\,dex.
The reason of such a large systematic offset for RCs is because that RCs (core helium burning) have higher stable luminosities (i.e. smaller values of log\,$g$) than those red giant branch stars (hereafter RGBs; shell hydrogen burning) with the same colors of RCs.
Identifying RCs from those RGBs with similar colors are very difficult, even with spectra available (see, e.g. Bovy et al. 2014; Huang et al. 2015a).
The stellar colors from SMSS can not offer enough sensitivities to distinguish the RCs and RGBs, and thus the significant offset for log\,$g \sim 2.3$ is apparent in our fit, as shown in Fig.\,4(b).
The dispersion of the fit residuals is about 0.3\,dex.

\subsection{Effective temperature}
We derive effective temperatures of red giants from the color $(V-K_{s})_0$, which is the best temperature indicator (see Huang et al. 2015b), combined with photometric metallicity given by Eq.\,(4), using the empirical metallicity-dependent effective temperature--color relation from Ram{\'{\i}}rez \& Mel{\'e}ndez (2005) or Huang et al. (2015b).
The effective temperature scale of Huang et al. (2015b) is based on interferometric data, and is only applicable for metal-rich stars of [Fe/H]\,$\ge -0.6$.
We thus adopt the relation of Ram{\'{\i}}rez \& Mel{\'e}ndez (2005) for stars of metallicity [Fe/H]\,$ < -0.6$, and use a temperature zero point scaled to that of Huang et al. (2015b)\footnote{The zero point difference between the temperature scales of Ram{\'{\i}}rez \& Mel{\'e}ndez (2005) and Huang et al. (2015b) is about $-16$\,K for giant stars according to Huang et al. (2015b).}. 
Here, $V$ band magnitudes are taken from APASS DR9 (Henden et al. 2016) and $K_{\rm s}$ band is taken from 2MASS (Skrutskie et al. 2006).
For stars without good photometry in APASS DR9, the color $g-K_s$ (with $g$ from SkyMapper) is used instead and is converted to $V-K_{s}$.
Using over half million red giants (selected from the SMSS DR1.1; see Section\,6 for details) with high quality photometry in APASS DR9 ($V$ band uncertainty smaller than 0.035\,mag), 2MASS ($K_{\rm s}$ band uncertainty smaller than 0.035\,mag) and SMSS DR1.1 ($g$ band uncertainty smaller than 0.035\,mag), we have obtained the convert relation: $V-K_{s} = 0.140 + 0.82\times(g-K_{s}) + 0.014\times(g-K_{s})^2$, with a scatter of 0.10\,mag.

\subsection{Distance}
To derive the distances of red giants, we apply two different approaches split by distance.
For nearby bright stars, we have accurate parallax measurements from Gaia DR2 (Lindegren et al. 2018)
and adopt distance estimates from Bailer-Jones et al. (2018).
For distant faint stars without good parallax measurements, we estimate absolute magnitudes from empirically calibrated color-luminosity fiducials using the measured intrinsic color $(g-i)_0$ and the photometric metallicity estimated above.
Photometric distances are then estimated by comparing the apparent magnitudes (after reddening correction) and the estimated absolute magnitudes. 
This approach for faint distant stars is very similar to the one developed by Xue et al. (2014; hereafter X14).
Here we briefly describe this method and refer to X14 for more technical details.

\begin{table*}
\centering
\caption{Comparisons with other spectroscopic samples.}
\begin{threeparttable}
\begin{tabular}{cccccccc}
\hline
Source& $\Delta T_{\rm eff}$&$\sigma T_{\rm eff}$&$\Delta$\,log\,$g$&$\sigma$\,log\,$g$&$\Delta$\,[Fe/H]&$\sigma$\,[Fe/H]&$N$\\
&(K)&(K)&(dex)&(dex)&(dex)&(dex)&\\
\hline
PASTEL&$-60$&86&0.48&0.40&$0.02$\tnote{a}&0.19\tnote{a}&186\\
GALAH DR2&$-84$&94&0.05&0.50&$-0.10$&0.17&38\,756\\
APOGEE DR14&$-80$&67&0.02&0.36&$-0.08$&0.18&3715\\
\hline
\end{tabular}
\begin{tablenotes}
\item[a] For stars of metallicity [Fe/H] greater than $-2.6$ only.
\end{tablenotes}
\end{threeparttable}
\end{table*}

\subsubsection{Color-Magnitude Fiducials}
Similar to X14, we use a set of giant-branch fiducials provided empirically by star clusters of metallicities ranging from [Fe/H]\,$= - 2.50$ to [Fe/H]\,$= + 0.50$, instead of giant branches from model isochrones.
To build the giant-branch cluster fiducials, we first adopt six cluster fiducial sequences (one open cluster and five globular clusters) derived from PS1 photometry (Bernard et al. 2014).
Colors and magnitudes of the PS1 photometric system are then converted to those of the SkyMapper system using the transformation equations from W18.
All six clusters have very accurate distance, reddening and metallicity determinations in the literature.
Our final adopted values of distance, reddening and metallicity are listed in Table\,2.
The resulting giant-branch fiducials, $M_{i_{0}}$ versus $(g-i)_0$, are shown in Fig.\,5. 

Again as in X14, we construct a denser homogeneously distributed grid of fiducials by interpolating the six cluster fiducials.    
Doing so, a quadratic interpolation function $c (M_{i_0}, {\rm [Fe/H]})$ is used to generate new fiducials,
where $c$ represents the color $(g-i)_0$.
In total, 31 new fiducials are interpolated in this way, in steps of 0.1 dex in metallicity from $-2.50$ to $+0.50$.
The new interpolated fiducials are also shown in Fig.\,5.

\subsubsection{Absolute Magnitude Likelihoods}
Using the interpolated fiducials, we then derive the absolute magnitude with a likelihood function of the form
\begin{equation}
\mathcal{L}_{X} = \frac{1}{\sqrt{2\pi}\sigma_{X_{\rm obs}}}\exp[\frac{-(X_{\rm obs} - X_{\rm model})^2}{2\sigma_{X_{\rm obs}}^2}] \text{,}  
\end{equation}
where $X_{\rm obs} = \{c, {\rm [Fe/H]}\}$ are assumed to be independent Gaussian observables and $c$ represents the color $(g-i)_0$. 
$X_{\rm model}$ represents the same quantity from the interpolated fiducials.
The combined likelihood is:
\begin{equation}
\mathcal{L} = \mathcal{L}_{\rm c}\mathcal{L}_{\rm [Fe/H]}  \text{.}
\end{equation}

For each star, Eq.\,(7) yields a probability distribution function (PDF) of $M_{i_0}$ and this PDF is used to derive the median and 68\% intervals of $M_{i_0}$.
Unlike in X14, here we do not introduce any priors (e.g. luminosity function and metallicity distribution of red giants) in the estimation.
This is because the knowledge of those priors is quite poor and we simply assume they are uniform.
Finally, distances of red giants are calculated by comparing the absolute magnitudes yielded by Eq.\,(7) and the reddening-corrected apparent magnitudes.

\subsubsection{Caveats}
We note the second approach based on the likelihood method for faint distant stars is only suitable for red giant branch stars.
Luminosities (absolute magnitudes) of RCs and red horizontal branch (HB) stars will be systematically underestimated by this method.
Fortunately, metal-rich (typically [Fe/H]\,$\ge -1.0$) RCs and relatively metal-rich ($-1.7 \leq$\,[Fe/H]\,$\leq -0.3$; Chen et al. 2011) red HB stars are mostly distributed close to the disk ($|Z| < 4$-$5$\,kpc) 
and therefore most of them have good distance estimates from Gaia parallax measurements.

\begin{figure*}
\begin{center}
\includegraphics[scale=0.425,angle=0]{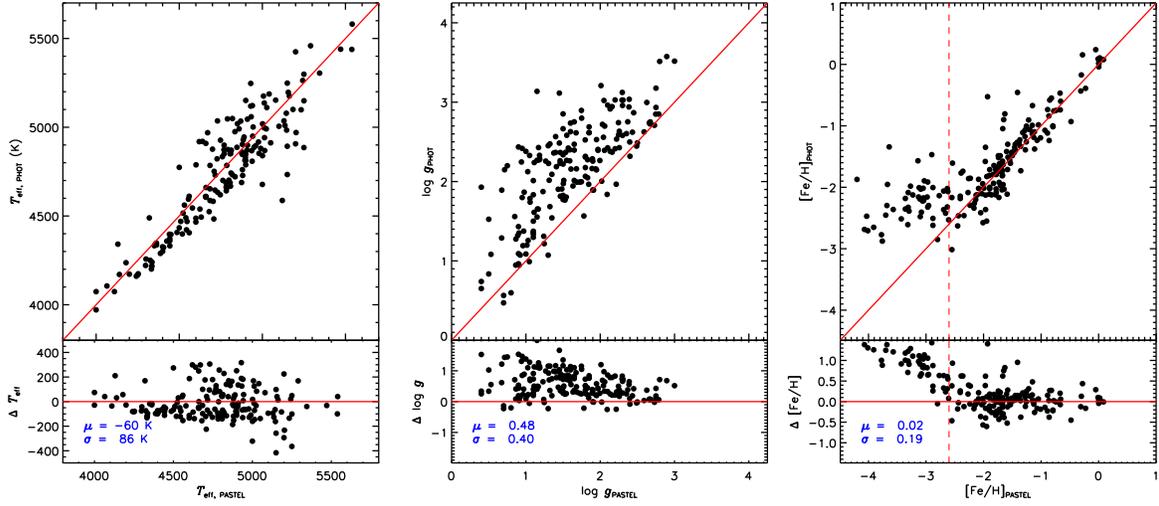}
\caption{Comparison of estimated atmospheric parameters with those from the PASTEL catalog.
              The differences are shown in the lower part of each panel, with the mean and standard deviation marked in the bottom-left corner.}
\end{center}
\end{figure*}

\begin{figure*}
\begin{center}
\includegraphics[scale=0.425,angle=0]{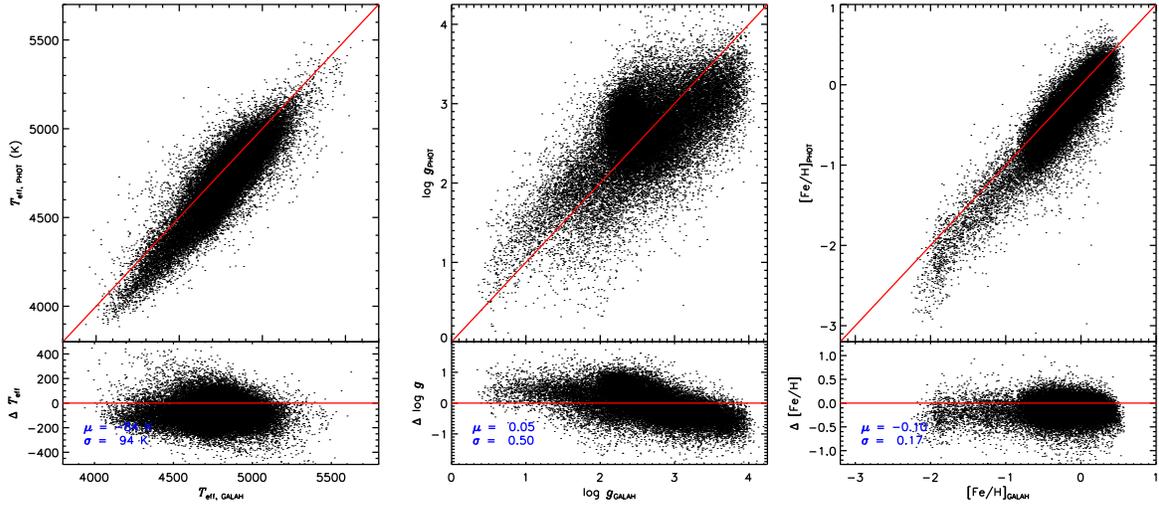}
\caption{Same as Fig.\,6 but for comparison with estimates from GALAH.}
\end{center}
\end{figure*}

\begin{figure*}
\begin{center}
\includegraphics[scale=0.425,angle=0]{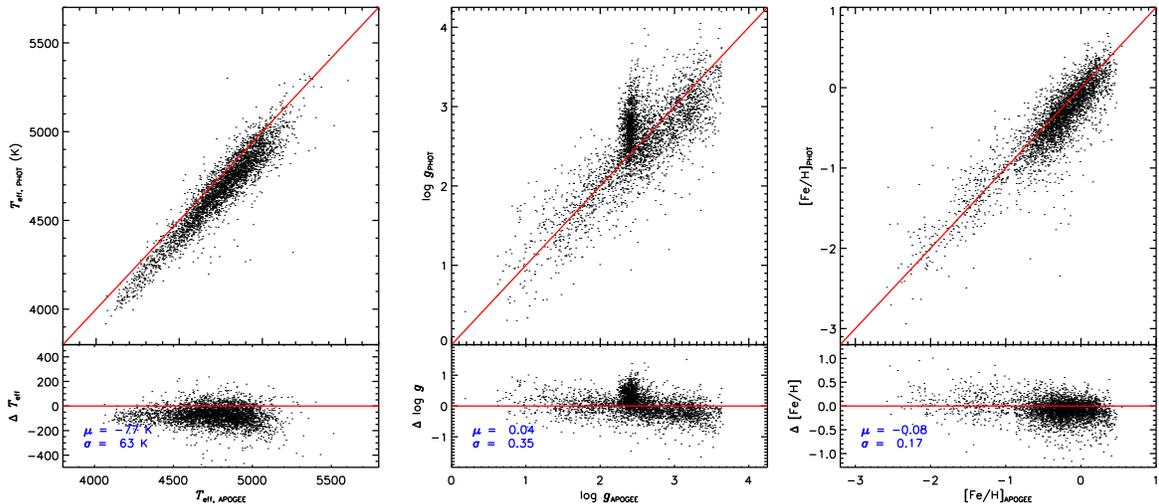}
\caption{Same as Fig.\,6 but for comparison with estimates from APOGEE.}
\end{center}
\end{figure*}

\begin{figure*}
\begin{center}
\includegraphics[scale=0.425,angle=0]{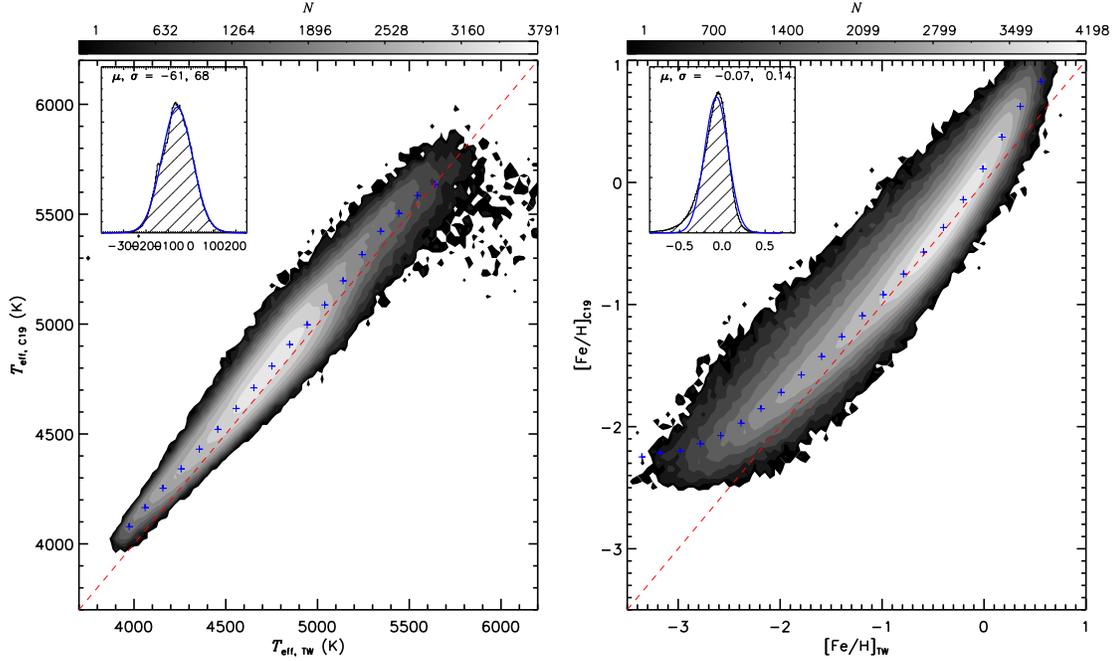}
\caption{Comparisons of estimated effective temperature $T_{\rm eff}$ (left panel) and metallicity [Fe/H] (right panel) with those derived by C19 for over 50\,0000 common stars.
In each panel, the color-coded contour of stellar number density in logarithmic scale is shown.
Crosses in blue are median values of our estimates calculated in bins of the values from this work.
In the left-top corner, the distribution of differences of $T_{\rm eff}$ and [Fe/H] between our estimates and the C19 values are shown.
Blue lines are Gaussian fits to the distribution, with the mean and dispersion of the Gaussian marked in the plot.}
\end{center}
\end{figure*}


\begin{figure*}
\begin{center}
\includegraphics[scale=0.35,angle=0]{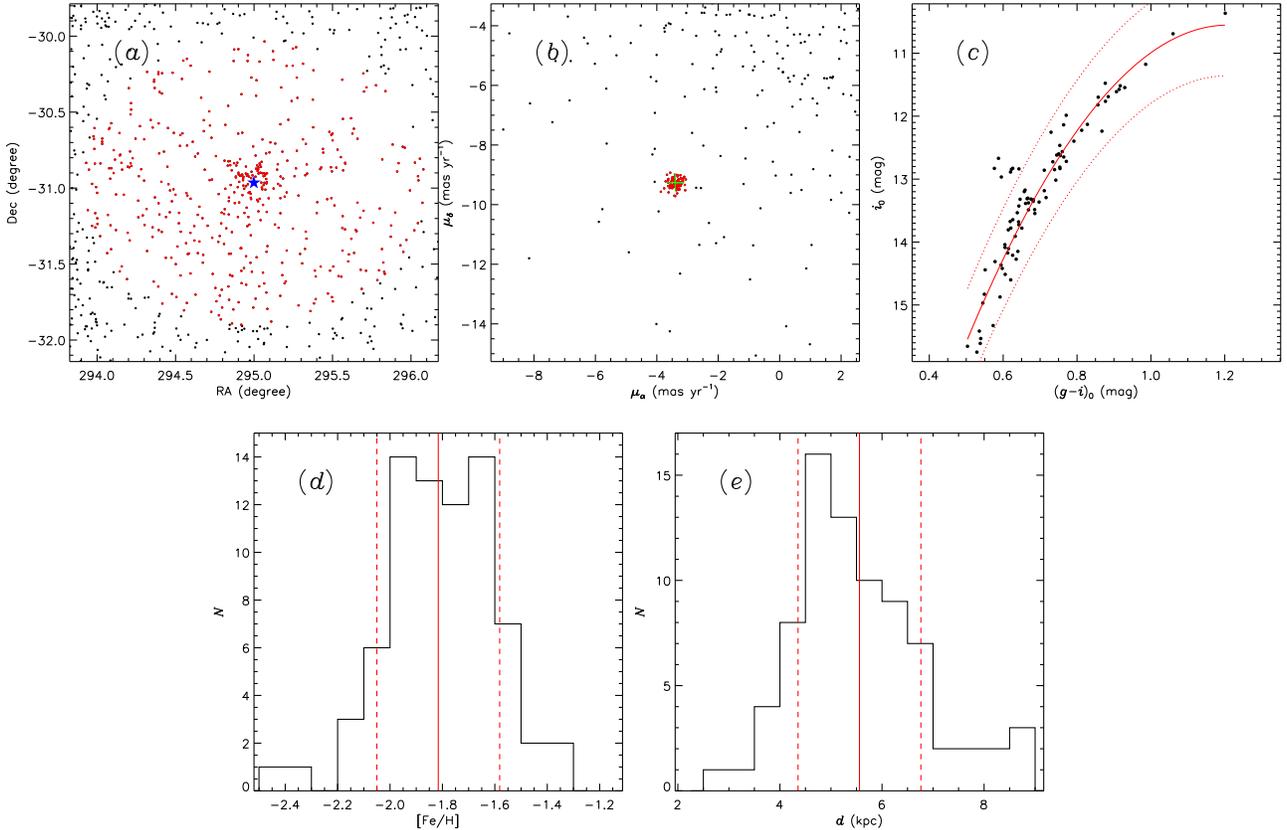}
\caption{Selection of member candidates of GC NGC\,6809, based on celestial coordinates (panel $a$), proper motions (panel $b$) and positions on CMD (panel $c$).
The blue star in panel ($a$) indicates the central position of NGC\,6809.
The dots in panel ($b$) are stars within 20 $r_{\rm h}$ from the central position, i.e. the red dots in panel ($a$).
The green plus in panel ($b$) indicates the proper motions of NGC\,6809.
The dots in panel (c) are stars  with $|\mu_{\alpha} - \mu_{\alpha, {\rm NGC\,6809}}| \leq 6$\,mas\,yr$^{-1}$ and\,$|\mu_{\delta} - \mu_{\delta, {\rm NGC\,6809}}| \leq 6$\,mas\,yr$^{-1}$, i.e. the red dots in panel ($b$).
The red line in panel ($c$) represents a second-order polynomial fit to the red giant branch and the two red dotted lines represent 2.5$\sigma$ deviations from the polynomial fit. 
Stars within the two red dotted lines are selected as the final member stars.
Panels ($d$) and ($e$) show distributions of metallicities and distances for the final sample of member stars.
The red line in each panel indicates the median value of the distribution and the red dashed lines mark $1$\,$\sigma$ dispersions from the median value.}
\end{center}
\end{figure*}

\begin{figure*}
\begin{center}
\includegraphics[scale=0.5,angle=0]{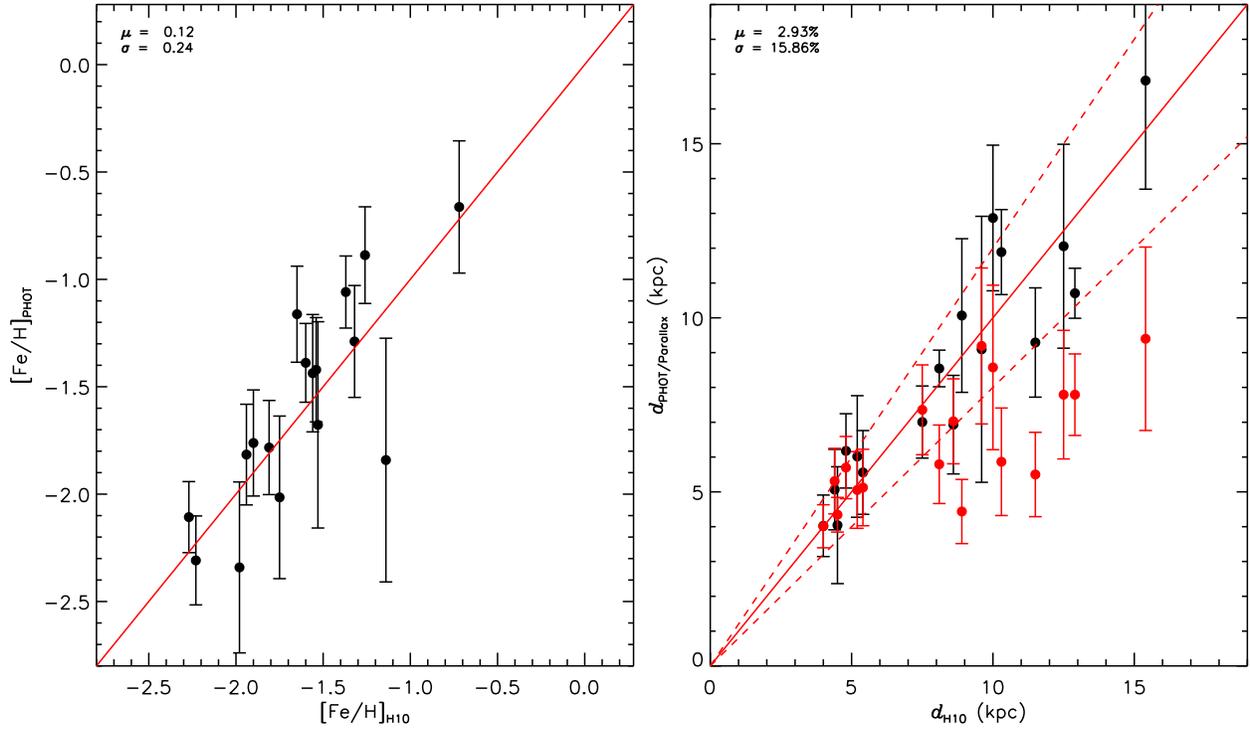}
\caption{Metallicities (left panel) and distances (right panel, black dots) of GCs derived with our photometric method compared to those from H10.
The median and dispersion of the (relative) differences are marked in the top-right corner of each panel. 
Distances of GCs (red dots) from the Gaia parallaxes are also overplotted in the right panel.
The two red dashed lines in the right panel mark $d_{\rm PHOT} = 1.2d_{\rm H10}$ and $d_{\rm PHOT} = 0.8d_{\rm H10}$, respectively.}
\end{center}
\end{figure*}

\begin{figure}
\begin{center}
\includegraphics[scale=0.4,angle=0]{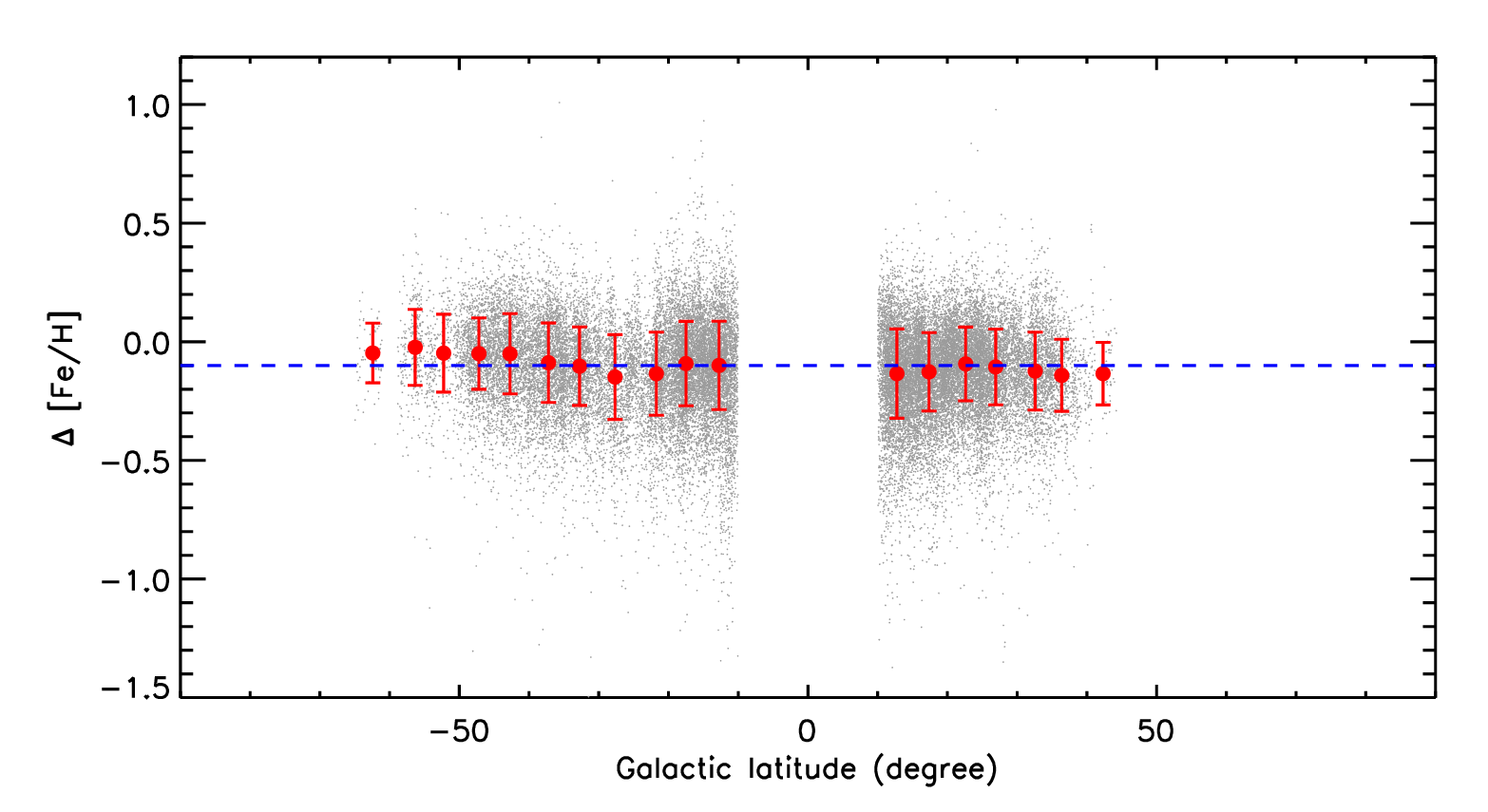}
\includegraphics[scale=0.4,angle=0]{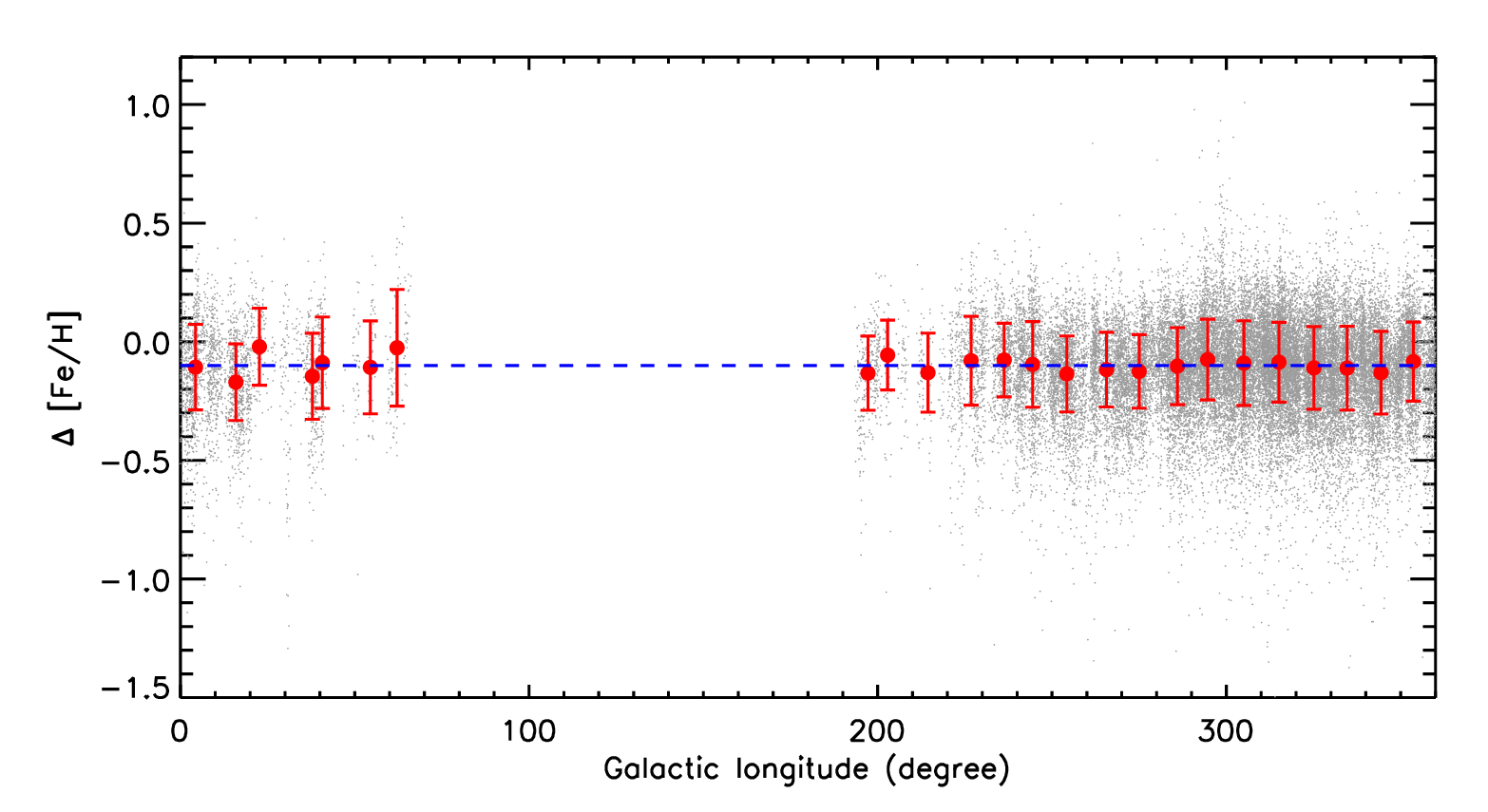}
\caption{Metallicity residuals (our photometric values minus GALAH spectroscopic ones) as a function of Galactic latitude (upper panel) and longitude (lower panel).
The red dots and error bars represent the means and standard deviations of the residuals in the individual Galactic latitude bins.
The blue dashed line marks a mean residual of $-0.1$ dex as found in Section\,5.1.}
\end{center}
\end{figure}

\section{Validation of atmospheric parameters and distance estimates}
\subsection{Comparison with spectroscopic samples}
We examine the accuracy of the atmospheric parameters estimated with our photometric methods by comparing our results with independent measurements from a number of spectroscopic samples, including the PASTEL catalog (Soubiran et al. 2016), the GALAH survey (Buder et al. 2018) and the APOGEE survey (Abolfathi et al. 2017).

PASTEL is a bibliographical compilation of measurements of stellar atmospheric parameters ($T_{\rm eff}$, log\,$g$, [Fe/H]) from different groups, mostly obtained by analyzing spectra with high resolution ($R \ge 30\,000$) and high SNR ($\ge 100$). 
As of the version of 2016 May, over 10\,000 stars catalogued by PASTEL have at least one set of measurements for all three parameters, $T_{\rm eff}$, log\,$g$ and [Fe/H]. 
We have cross-matched our red giant star sample with those $\sim 10\,000$ stars in the PASTEL catalogue that have all three basic stellar atmospheric parameters\footnote{For stars with more than one estimate for a given parameter, mean values (after 3$\sigma$ clipping) are adopted.} available and found 186 stars in common.
The comparisons are shown in Fig.\,6.
For effective temperature $T_{\rm eff}$, our photometric results agrees well with the measurements from PASTEL for $T_{\rm eff, PASTEL} \ge 4700$\,K, but are significantly lower than the PASTEL results by $\sim 100$\,K at lower temperatures.
We note this discrepancy does not necessarily imply any problems of our temperature scale considering that PASTEL is simply a compilation of the literature data that are highly inhomogeneous in nature.
The dispersion in the difference between our and PASTEL measurements of $T_{\rm eff}$ is quite small with only 86\,K.
For surface gravity, our photometric results are systematically higher than those of PASTEL by 0.48\,dex, along with a dispersion of 0.4\,dex. 
The cause of this large systematic and uncertainty is unclear.
One possibility is calibration issues in the SkyMapper $u$ band photometry as discussed in Casagrande et al. (2018).
Another possibility is caused by our photometric calibration.
Finally, for metallicity, our photometric results are in excellent agreement with measurements from high resolution spectroscopy collected by PASTEL, at metallicity [Fe/H]$_{\rm PASTEL}$\,$\ge -2.6$.
The differences between our and PASTEL values for [Fe/H]$_{\rm PASTEL}$\,$\ge -2.6$  have a mean of only 0.02\,dex, along with a dispersion of 0.19\,dex.
Our photometric metallicities deviate from the high resolution values for [Fe/H]$_{\rm PASTEL}$\, $< -2.6$.
This is easy to understand considering that our calibration presented in Section\,4.1 is only applicable for [Fe/H]$_{\rm PASTEL}$\,$\ge -2.6$ and the fact that SkyMapper colors become less sensitive to metallicity for [Fe/H]\,$< - 2.6$.

The GALAH survey aims to collect spectra for $\sim$\,one million stars with the fiber-fed high-resolution ($R =$\,28\,000) spectrograph HERMES that covers four discrete optical wavelength ranges, $4713$--$4903$\AA, $5648$--$5873$\AA, $6478$--$6737$\AA, and $7585$--$7887$\AA.
The spectrograph is mounted on the 3.9-meter Anglo-Australian Telescope (AAT) at Siding Spring Observatory (De Silva et al. 2015). 
The recently released GALAH DR2 (Buder et sl. 2018) contains measurements of stellar atmospheric parameters and abundances of 23 elements for 342\,682 stars. 
We have cross-matched our red giant star sample with the GALAH DR2, and found 38\,756 common stars with good parameter estimates (i.e. {\it flag\_cannon}\,=\,0).
A comparison of our photometric atmospheric parameters and those from GALAH DR2 is shown in Fig.\,7.
Our photometric temperatures agree well with the GALAH results, with a mean difference of $-84$\,K and a dispersion of $94$\,K.
For surface gravity, some systematic differences between our photometric and the GALAH results as a function of the GALAH log\,$g$ are apparent. 
We believe that this is largely due to the systematics of our calibration presented in Section\,4.2.
The relatively large dispersion (around 0.5 dex) of the differences is contributed by uncertainties in both measurements, which are greater than $\sim$\,0.3\,dex in both data sets (Buder et al. 2018).
The large systematic offset, due to the issue of our photometric relation (see Section\,4.2 for details), around log\,$g \sim 2.3$ is also seen in Fig.\,7.
For metallicity, our photometric results agree very well with the GALAH values, with a mean difference of 0.1\,dex and a dispersion of 0.17\,dex.
This 0.1\,dex offset is likely a problem of the GALAH measurements, considering that our photometric results match those of PASTEL within 0.02\,dex.
Buder et al. (2018) also report that GALAH might have overestimated [Fe/H] by as large as 0.35\,dex for cool giants when compared to the Gaia benchmark stars (Jofr{\'e} et al. 2014).
In this context, Buder et al. (2018) mention that GALAH may also have overestimated the metallicities of metal-poor stars (i.e. [Fe/H]\,$\leq -1.8$\,dex) by 0.5 to 1 dex.

\begin{table*}
\centering
\caption{ Comparison of photometric metallicities and distances with the values from H10 for GCs}
\begin{threeparttable}
\begin{tabular}{cccccccccc}
\hline
Name&$l$&$b$&$\mu_{\alpha}$&$\mu_{\delta}$&$d_{\rm H10}$&[Fe/H]$_{\rm H10}$&$n$& $d_{\rm PHOT}$&[Fe/H]$_{\rm PHOT}$\\
&(degree)&(degree)&(mas\,yr$^{-1}$)&(mas\,yr$^{-1}$)&(kpc)&&&(kpc)&\\
\hline
 NGC 0104&305.89&$-44.89$&$  5.2477\pm  0.0016$&$ -2.5189\pm  0.0015$&$  4.5$&$ -0.7$& 41&$  4.0\pm  1.7$&$ -0.7\pm  0.3$\\
 NGC 0288&152.30&$-89.38$&$  4.2385\pm  0.0035$&$ -5.6470\pm  0.0026$&$  8.9$&$ -1.3$& 13&$ 10.1\pm  2.2$&$ -1.3\pm  0.3$\\
 NGC 0362&301.53&$-46.25$&$  6.6954\pm  0.0045$&$ -2.5184\pm  0.0034$&$  8.6$&$ -1.3$& 30&$  6.9\pm  1.4$&$ -0.9\pm  0.2$\\
 NGC 1904&227.23&$-29.35$&$  2.4702\pm  0.0048$&$ -1.5603\pm  0.0054$&$ 12.9$&$ -1.6$& 14&$ 10.7\pm  0.7$&$ -1.4\pm  0.2$\\
 NGC 2808&282.19&$-11.25$&$  1.0032\pm  0.0032$&$  0.2785\pm  0.0032$&$  9.6$&$ -1.1$& 19&$  9.1\pm  3.8$&$ -1.8\pm  0.6$\\
 NGC 4590&299.63&$ 36.05$&$ -2.7640\pm  0.0050$&$  1.7916\pm  0.0039$&$ 10.3$&$ -2.2$& 44&$ 11.9\pm  1.2$&$ -2.3\pm  0.2$\\
 NGC 5139&309.10&$ 14.97$&$ -3.1925\pm  0.0022$&$ -6.7455\pm  0.0019$&$  5.2$&$ -1.5$&164&$  6.0\pm  1.7$&$ -1.7\pm  0.5$\\
 NGC 5897&342.95&$ 30.29$&$ -5.4108\pm  0.0053$&$ -3.4595\pm  0.0045$&$ 12.5$&$ -1.9$& 12&$ 12.1\pm  2.9$&$ -1.8\pm  0.2$\\
 NGC 6093&352.67&$ 19.46$&$ -2.9469\pm  0.0090$&$ -5.5613\pm  0.0073$&$ 10.0$&$ -1.8$& 12&$ 12.9\pm  2.1$&$ -2.0\pm  0.4$\\
 NGC 6101&317.74&$-15.82$&$  1.7500\pm  0.0060$&$ -0.4000\pm  0.0060$&$ 15.4$&$ -2.0$& 12&$ 16.8\pm  3.1$&$ -2.3\pm  0.4$\\
 NGC 6218& 15.72&$ 26.31$&$ -0.1577\pm  0.0040$&$ -6.7683\pm  0.0027$&$  4.8$&$ -1.4$& 20&$  6.2\pm  1.1$&$ -1.1\pm  0.2$\\
 NGC 6254& 15.14&$ 23.08$&$ -4.7031\pm  0.0039$&$ -6.5285\pm  0.0027$&$  4.4$&$ -1.6$& 41&$  5.1\pm  1.2$&$ -1.4\pm  0.3$\\
 NGC 6541&349.29&$-11.19$&$  0.2762\pm  0.0054$&$ -8.7659\pm  0.0048$&$  7.5$&$ -1.8$& 11&$  7.0\pm  1.0$&$ -1.8\pm  0.2$\\
 NGC 6752&336.49&$-25.63$&$ -3.1908\pm  0.0018$&$ -4.0347\pm  0.0020$&$  4.0$&$ -1.5$& 64&$  4.0\pm  0.9$&$ -1.4\pm  0.2$\\
 NGC 6809&  8.79&$-23.27$&$ -3.4017\pm  0.0031$&$ -9.2642\pm  0.0028$&$  5.4$&$ -1.9$& 79&$  5.6\pm  1.2$&$ -1.8\pm  0.2$\\
 NGC 7089& 53.37&$-35.77$&$  3.4911\pm  0.0077$&$ -2.1501\pm  0.0071$&$ 11.5$&$ -1.6$& 32&$  9.3\pm  1.6$&$ -1.2\pm  0.2$\\
 NGC 7099& 27.18&$-46.84$&$ -0.7017\pm  0.0063$&$ -7.2218\pm  0.0055$&$  8.1$&$ -2.3$& 25&$  8.5\pm  0.5$&$ -2.1\pm  0.2$\\
\hline
\end{tabular}
\begin{tablenotes}
\item[] Col.\,1 gives the cluster identification number, Cols. 2 and 3 give the central Galactic longitudes and latitudes of the clusters from H10, Cols. 4 and 5 present the proper motions of the GCs from Gaia Collaboration (2018b), Cols. 6 and 7 give the distances and metallicities of the GCs from H10, Col. 8 gives the number of GC member stars that pass the cuts presented in Section\,5.2. Cols. 9 and 10 present the photometric metallicities and distances of the GCs (see Section\,5.2).
\end{tablenotes}
\end{threeparttable}
\end{table*}

The APOGEE survey aims to collect spectra of high resolution ($R \sim $22\,500) and high SNR ($\sim$100 per pixel) in the near-infrared ($H$ band, 1.51-1.70\,$\mu$m) mainly red giants in the Milky Way.
More details of the survey, including scientific motivation, target selection, data reduction and stellar parameter determination, can be found in Zasowski et al. (2013), Majewski et al. (2017), Nidever et al. (2015) and Garc{\'{\i}}a P{\'e}rez et al. (2016).
The latest APOGEE DR14 has released measurements of stellar atmospheric parameters and elemental abundances for 263\,444 stars (Abolfathi et al. 2018).
We have cross-matched our sample of red giants and the APOGEE DR14, and found 3715 common stars with APOGEE parameters determined from high quality spectra (SNRs\,$\ge$\,100).
The comparisons between our photometric atmospheric parameters and those of the APOGEE DR14 are then shown in Fig.\,8.
Except for an offset of $-77$\,K, our photometric values of $T_{\rm eff}$ are well consistent with the APOGEE results, with a small dispersion of 63\,K.
The offset is largely due to the fact that the APOGEE $T_{\rm eff}$ is calibrated by the temperature scale from the infrared flux method of Gonz{\'a}lez Hern{\'a}ndez \& Bonifacio (2009), which is hotter than the interferometric scale by 50 to 100\,K (Huang et al. 2015b).
Our values of photometric surface gravity agree with the APOGEE results quite well, with a dispersion of 0.35\,dex (mainly contributed by the uncertainties of our photometric measurements).
The large deviations seen for RCs with log\,$g$ around 2.3 are caused by the issue  in our calibration relation (see Section\,4.2).
Finally, our photometric metallicities are in excellent agreement with the APOGEE results, with a mean offset of $-0.08$\,dex and a dispersion of $0.17$\,dex.

All the comparison results are presented in Table\,3.
In summary, our photometric stellar atmospheric parameters agree well in general with the currently available results from high-resolution spectroscopy.
For effective temperature, except the different temperature scales adopted, our photometric results are consistent with the high-resolution spectroscopic ones within 90\,K.
For surface gravity, except for the systematic trends from the calibration relation, our photometric results are otherwise well consistent with high-resolution spectroscopic ones within 0.30-0.35\,dex.
For metallicity, our photometric results are consistent with the high-resolution spectroscopic ones within 0.17-0.19\,dex.
Our photometric metallicity scale is in excellent agreement with the PASTEL (except for the most metal-poor stars of [Fe/H]\,$\le -2.6$), whose metallicities are all derived from spectra of resolution greater than 30\,000. 

As mentioned in Section\,1, C19 have also derived values of effective temperature $T_{\rm eff}$ using the infrared flux method (IRFM; e.g. Casagrande et al. 2010) and of metallicity [Fe/H] based on colors from the SMSS DR1.1 and 2MASS. 
We therefore compare our estimated $T_{\rm eff}$ and [Fe/H] with those  derived by C19 for over 50\,0000 common stars.
The results are presented in Fig.\,9.
Generally, our estimates are in excellent agreement with those of C19, except the offsets.
The offset of $-61$\,K (this work minus C19) on $T_{\rm eff}$ is because that the $T_{\rm eff}$ scale  of IRFM (Casagrande et al. 2010) adopted by C19 is about 100\,K hotter than the scale of Huang et al. (2015b) adopted by this work.
The photometric metallicity [Fe/H] of C19 is derived based on the relation trained by the SkyMapper-GALAH common stars.
As shown in Fig.\,7, the metallicity derived from the GALAH survey is about 0.1\,dex larger than our photometric estimates, systematically.
This is the reason that a clear offset of $-0.07$\,dex (this work minus C19) seen in Fig.\,9.
The larger deviations for metal-poor stars ([Fe/H]\,$\leq -2.0$) is because that the photometric relation is restricted with [Fe/H]\,$> -2.0$.

\begin{figure}
\begin{center}
\includegraphics[scale=0.475,angle=0]{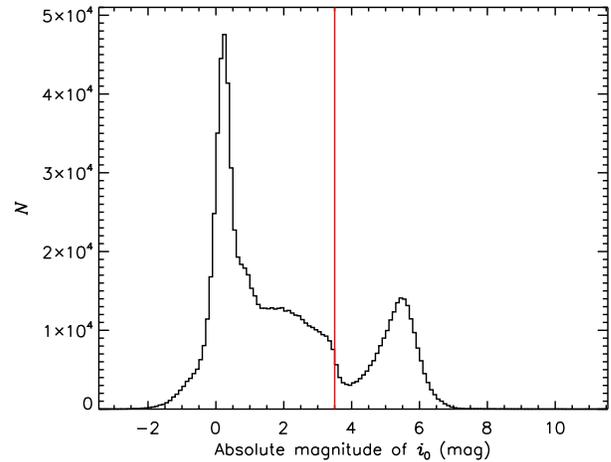}
\caption{Histogram of $M_{i_{0}}$ of 897,867 stars that pass the parallax quality cut ($\sigma_{\varpi}$/$\varpi \le 0.2$) and the distance cut (smaller than 4.5\,kpc) in the sample of 1,188,707 red giant star candidates selected from the SMSS DR1.1.}
\end{center}
\end{figure}

\begin{figure*}
\begin{center}
\includegraphics[scale=0.45,angle=0]{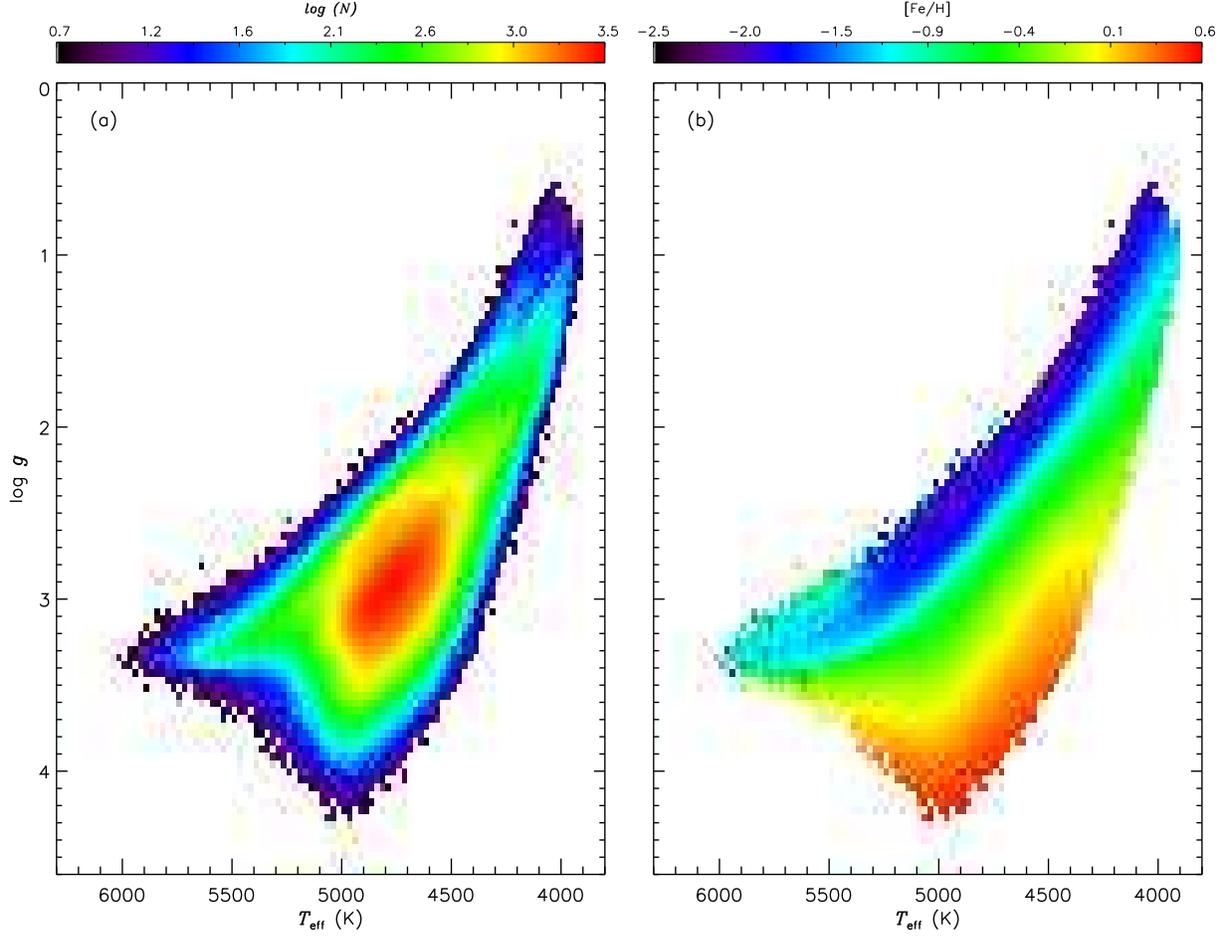}
\caption{Number density (left, in logarithmic scale) and metallicity (right) distributions of the sample of nearly one million red giants in the $T_{\rm eff}$--log\,$g$ plane.
              The number densities (for a bin size of 25\,K in $T_{\rm eff}$ and 0.05\,dex in log\,$g$) are indicated by the top-left color bar.
              The median values of metallicity in a given bin (of the same bin sizes as for the number densities) are indicated by the top-right color bar. 
              All bins shown in the plots have at least 5 stars.}
\end{center}
\end{figure*}

\begin{figure}
\begin{center}
\includegraphics[scale=0.425,angle=0]{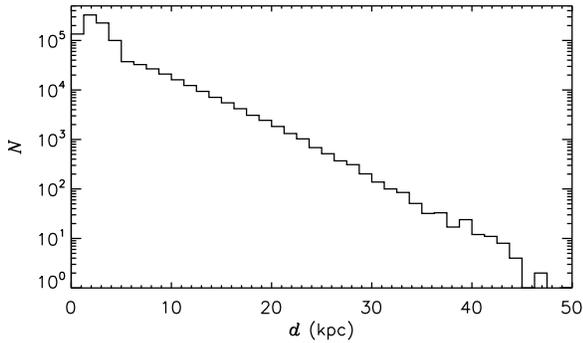}
\caption{Distance distribution of the final red giant sample.}
\end{center}
\end{figure}

\begin{figure*}
\begin{center}
\includegraphics[scale=0.45,angle=0]{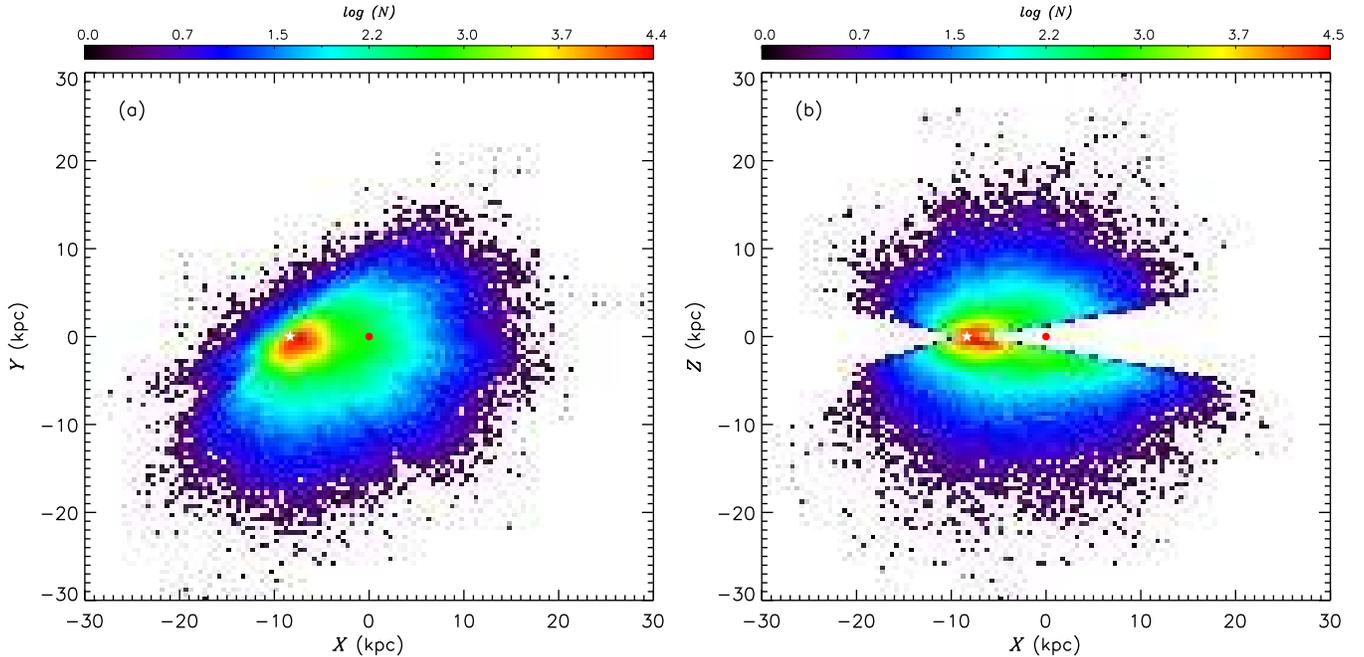}
\caption{Spatial distributions of our final sample of nearly one million red giants in the $X$-$Y$ (left) and $X$-$Z$ (right) planes. 
              The Sun is located at ($X$, $Y$, $Z$) = ($-8.34$,\,$0.0$,\,$0.0$)\,kpc.
              The stellar number densities (for a bin size of 0.5 kpc in both axes) are indicated by the top colorbars.
              In each panel, the white star and the red dot indicate the position of the Sun and the Galactic center, respectively.}
\end{center}
\end{figure*}

\begin{figure}
\begin{center}
\includegraphics[scale=0.35,angle=0]{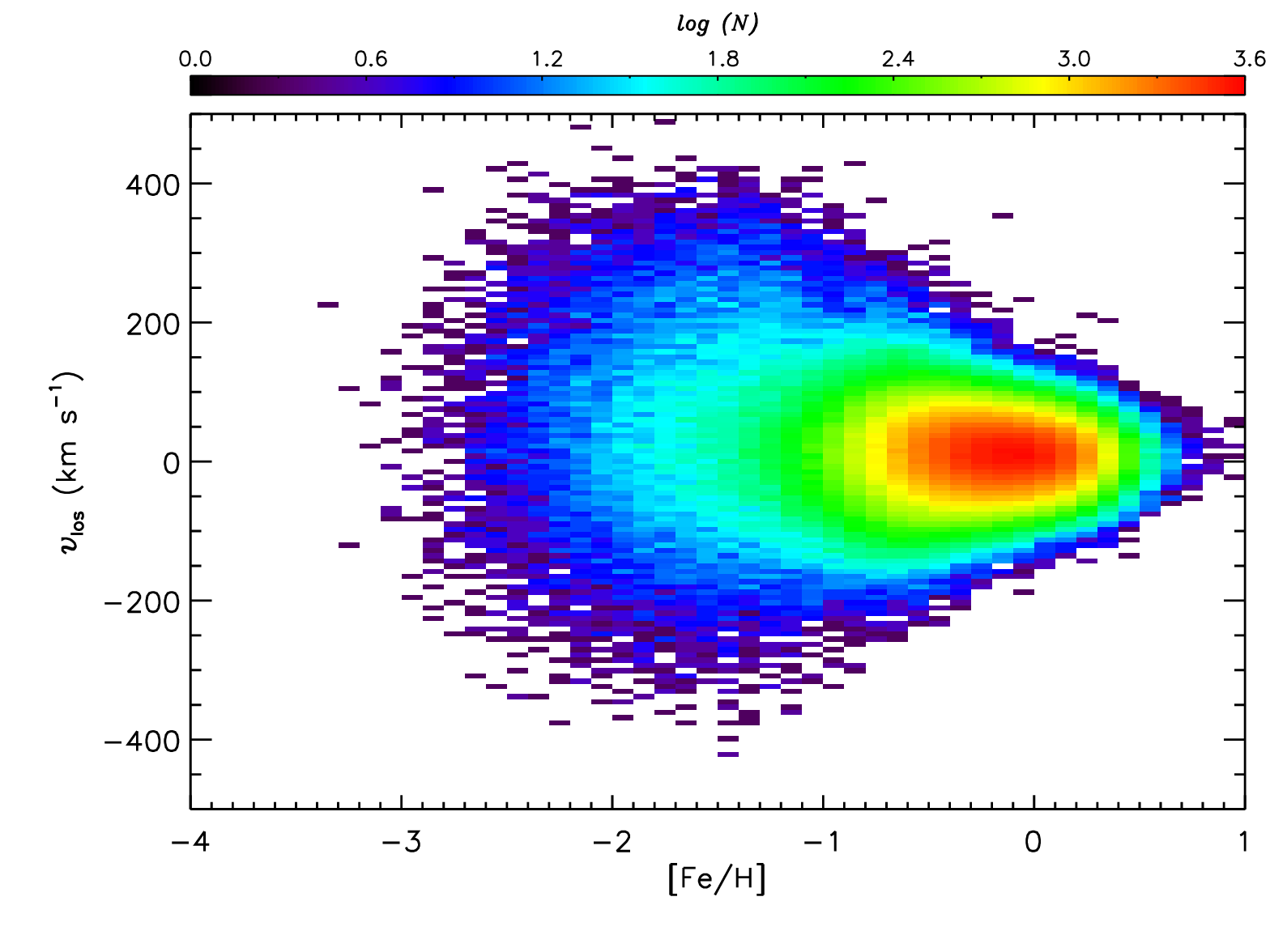}
\caption{Number distribution (on a logarithmic scale) of stars with radial velocity measurements available from the current large-scale Galactic spectroscopic surveys in the [Fe/H]-$v_{\rm los}$ panel.
                            The number densities (for a bin size of 0.1\,dex in [Fe/H] and 7.5 km\,s$^{-1}$ in $v_{\rm los}$) are indicated by the top color bar.}
\end{center}
\end{figure}

\subsection{Validation with globular clusters}
Stars of a given globular cluster (GC) are generally believed to form almost at the same time and location with the same metallicity.
Member stars of GCs thus serve as a good testbed to check the accuracy of metallicity and distance determinations.

In the SMSS DR1.1, dozens of GCs have been observed.
We select member stars of those GCs based on sky position, proper motion and position in the color-magnitude diagram (CMD), requiring that they:
\begin{itemize}[leftmargin=*]

\item are within 20 half-light radii ($r_{\rm h}$) from the center of the GC;

\item have proper motions of $|\mu_{\alpha} - \mu_{\alpha, {\rm GC}}| \leq 6$\,mas\,yr$^{-1}$ and\,$|\mu_{\delta} - \mu_{\delta, {\rm GC}}| \leq 6$\,mas\,yr$^{-1}$;

\item fall on the red giant branch, delineated by a second-order polynomial on the CMD.

\end{itemize}
Here $r_{\rm h}$ is from Harris et al. (2010; hereafter H10) and the proper motions of the GCs ($\mu_{\alpha, {\rm GC}}$ and $\mu_{\delta, {\rm GC}}$, see Table\,4) are taken from Gaia Collaboration (2018b). 
Fig.\,10 illustrates the process of selecting member candidates of NGC\,6809 as an example.
Of the three cuts, the second one is of vital importance.
As shown in Fig.\,10(b), the field and GC member stars are clearly separated in the $\mu_{\alpha}$-$\mu_{\delta}$ diagram, owing to the highly accurate proper motion measurements from Gaia DR2.
In Fig.\,10(d-e), we show the metallicity and distance distribution of the final sample of potential member stars of NGC\,6809.
For each GC, we then estimate the median metallicity and distance as well as uncertainties from the distribution of those quantities in the sample.
In total, mean photometric metallicities and distances of 17 GCs are derived in this way and the results are presented in Table\,4.

Compared to values of H10, our photometric results differ by an insignificant mean offset of $0.12$\,dex and a dispersion of $0.24$\,dex (see Fig.\,11).
For distance, our photometric results match H10 ones with a negligible mean offset, that translates into an average distance difference ($\frac{\Delta d}{d}$) of 3 per cent and a dispersion of 16 per cent (see Fig.\,11). 
The dispersion of the relative distance differences is consistent with the median precision found by X14 when using a similar technique.
We have also compared the distances from the Gaia parallaxes with those of H10.
Clearly, Gaia distances only work well for nearby GCs ($d \leq$\,4-6\,kpc) and
for those the Gaia distances are in excellent agreement with H10 and our photometric values.
We therefore believe that it is reasonable and self-consistent to combine the two sets of distances estimates in our work -- those from Gaia parallaxes for nearby stars and those from the photometric calibration for distant stars.

Finally, we note the recent work of Casagrande et al. (2018) on the determinations of the $uvgriz$ photometric zero-points of the SMSS DR1.1.
They found that the zero-points depend on Galactic latitude (especially for $|b| \leq 10^{\circ}$).
In the current work, we have not corrected for those potential variations in the photometric zero-points as they are found based on a small sample of bright stars (544 stars), that mostly have only $uv$ photometry, as they are brighter than 12\,mag in the $uv$ bands and saturated in the other bands.
It is true that our results could be potentially affected by such zero-point variations with Galactic latitude, but the effects are likely to be minor since our sample does not contain stars of $|b| \leq 10^{\circ}$. 
To examine the possible effects of those potential zero-point variations on the metallicity determination, we plot the metallicity residuals (our photometric values minus GALAH spectroscopic ones) as a function of Galactic latitude in Fig.\,12.
The mean residuals indeed show some small variations with Galactic latitude, but only at a level of 0.03-0.05\,dex.
In addition, the metallicity residuals as a function of Galactic longitude is also shown in Fig.\,12 and no evident trend is detected.

\section{The SMSS giant sample}

\subsection{Sample construction}
In this section, we attempt to construct a sample of red giants with estimates of atmospheric parameters and distances from SMSS DR1.1.
We first select stars ({CLASS\_STAR}\,$\ge$\,0.6) with good photometry, i.e. FLAGS\,$ = 0$, from SMSS DR1.1. 
We also require that the uncertainties in $uvgi$ magnitudes are smaller than 0.05\,mag such that the resulting colors are accurate enough to deliver robust estimates of atmospheric parameters and distances.
In addition, we exclude stars of $|b| < 10^{\circ}$, since most of those stars do not have magnitudes in $uv$ bands.
With those cuts, around 11 million stars are selected.
The giant star selection criteria defined by Eqs. (1)-(3) are then applied to these 11 million stars. 
This yields 1,188,707 potential red giants. 
To exclude potential contamination from dwarf stars, we show the number distribution of $M_{i_0}$ of 897,867 stars with good-quality parallax measurements (i.e. $\sigma_{\varpi}$/$\varpi \le 0.2$) and distances smaller than 4.5\,kpc in Fig.\,13.
Significant contamination from dwarf stars is clearly seen for $M_{i_0}$ between 3.5 and 8\,mag. 
We then exclude 215,713 potential contaminators by applying a cut of $M_{i_0} \ge 3.5$.
Finally, we have 972,994 (close to one million) highly probable red giants left.

\subsection{Sample content}

For this sample of nearly one million highly probable red giants, we derive their atmospheric parameters and distances from the SkyMapper colors using the methods described in Section 4.
The resulting number density and metallicity distributions of the sample in the $T_{\rm eff}$--log\,$g$ plane are shown in Fig.\,14. 
Most of the giant stars have $T_{\rm eff}$ around 4800\,K and log\,$g$ around 3.
The presence of a small fraction of metal-rich stars of log\,$g$ greater than 3.5 is caused by the systematic errors in the calibrated relation that is described in Section\,4.2.
As Fig.\,14(b) shows, the stars are bluer and brighter (smaller log\,$g$) when they are more metal-poor, consistent with stellar evolution theory.
For distances, we have directly adopted the results from Bailer-Jones et al. (2018) for stars with relative parallax uncertainties smaller than 20 per cent and distances smaller than 4.5\,kpc.
A total of 683,172 stars get their distances this way.
For the remaining 289,822 (distant) stars, we estimate their distances using the method described in Section\,4.4.
The distance distribution of the whole sample is presented in Fig.\,15. 
Most of the sample stars are within 10\,kpc, although some of them reach out as far as 40\,kpc.
From the distances, we have calculated the 3D positions of these stars in a right-handed Cartesian system ($X$,\,$Y$,\,$Z$), positioned around the Galactic center (GC)
with $X$ pointing in the direction opposite to the Sun, $Y$ in the direction of Galactic rotation and $Z$ towards the North Galactic Pole.
The Sun is assumed to be at ($X$, $Y$, $Z$) = ($-8.34$,\,$0.0$,\,$0.0$)\,kpc (Reid et al. 2014).
The resulting spatial distributions in $X$-$Y$ and $X$-$Z$ planes are shown in Fig.\,16.

\begin{table}
\centering
\caption{Sources of radial velocity measurements}
\begin{tabular}{ccc}
\hline
Source& Resolution& $N$\\
\hline
GALAH DR2&28,000&45\,512\\
APOGEE DR14&22,500&4698\\
Gaia DR2&11,500&362\,754\\
RAVE DR5 &7500&3338\\
LAMOST DR3&1800&7295\\
SEGUE DR9&1800&398\\
In total&--&423\,995\\
\hline
\end{tabular}
\end{table}

\begin{figure*}
\begin{center}
\includegraphics[scale=0.425,angle=0]{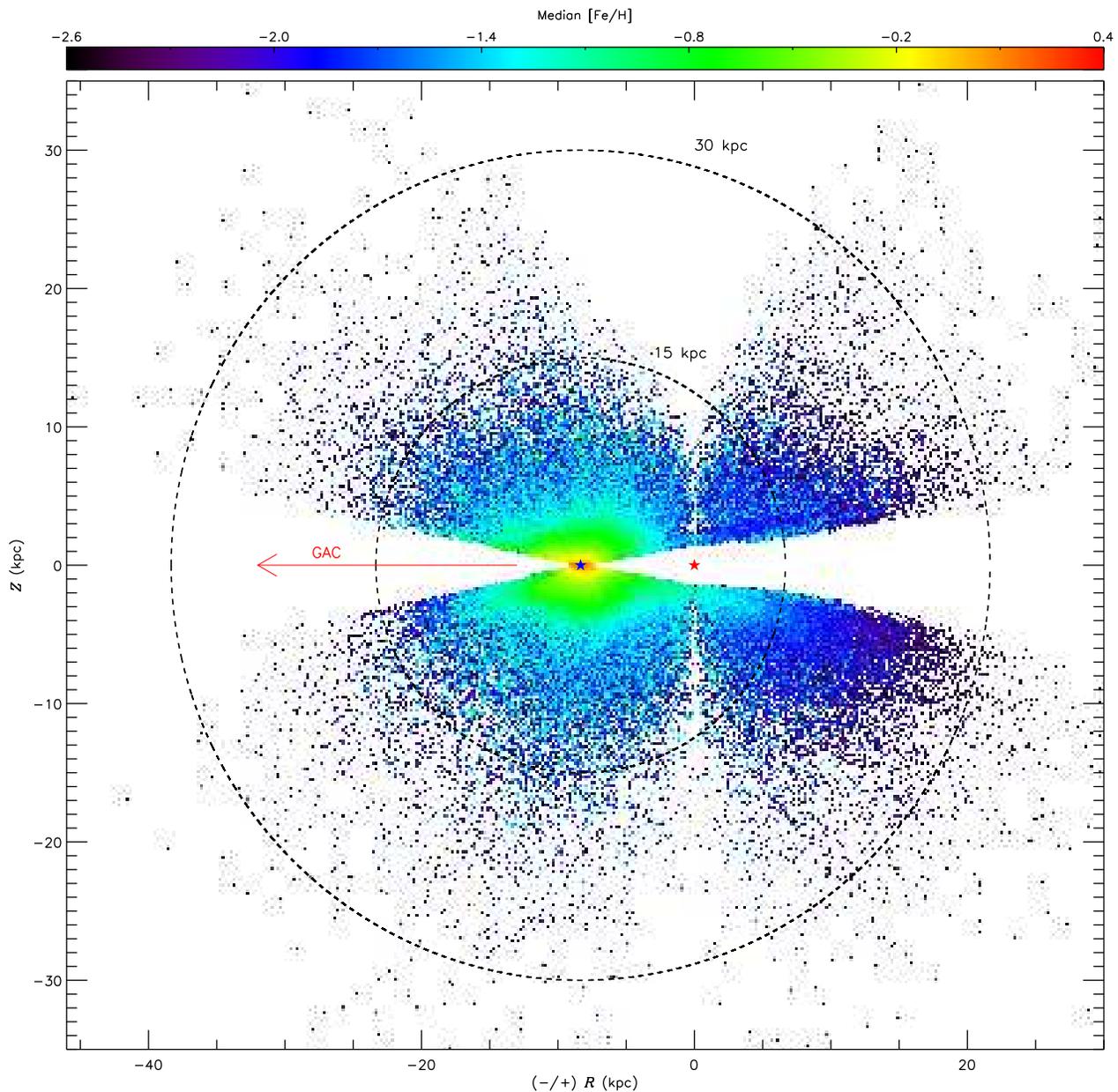}
\caption{Median metallicity [Fe/H] distribution in the $\pm R$ and $z$ plane from our red giant sample, binned by 0.20$\times$0.20\,kpc$^2$ in $\pm R$ and $z$.
Here, we adopt a negative value for $R$ (namely $-R$) if $X < 0$ and keep the original value of $R$ if $X \geq 0$.
The blue and red star mark the positions of the Sun and the Galactic center, respectively.
The two dashed circles mark distances of $15$ and $30$\,kpc from the Sun.}
\end{center}
\end{figure*}

In addition to atmospheric parameters and distances, we have also included proper motion and the radial velocity measurements of the sample stars.
Nearly all stars in the sample have accurate proper motion measurements from the Gaia DR2 (Lindegren et al. 2018).
For radial velocities, we take measurements from GALAH DR2 (Buder et al. 2018), SDSS/APOGEE DR14 (Abolfathi et al. 2017), Gaia DR2 (Katz et al. 2018), RAVE DR5 (Kunder et al. 2017), LAMOST DR3 (Xiang et al., 2017a; Huang et al. in prep.) and SDSS/SEGUE DR12 (Alam et al. 2012).
If a star has been targeted by more than two different surveys, we adopt the measurement from the higher-resolution survey.
The zero points of radial velocities yielded by different surveys are all calibrated to that given by the APOGEE radial velocity standard stars (Huang et al. 2018).
Radial velocity measurements for a total of 423,995 stars are obtained from those surveys. 
The actual numbers from the individual surveys are presented in Table\,5.
We show the stellar number distribution in the [Fe/H]-$v_{\rm los}$ plane in Fig.\,17.
A clear trend of decreasing radial velocity dispersion with [Fe/H] is seen in the plot, which 
is an indication of the robustness of our [Fe/H] determinations.
Finally, we derive 3D velocities for stars with radial velocity measurements in Cartesian, Galactocentric cylindrical and Galactocentric spherical systems.
The three velocity components are represented by ($U$,\,$V$,\,$W$) in the Cartesian system (see above) centered on the Sun, ($v_{R}$,\,$v_{\phi}$,\,$v_z$) in the Galactocentric cylindrical system and ($v_{r}$,\,$v_{\theta}$,\,$v_{\phi}$) in the Galactocentric spherical system.
In the Galactocentric cylindrical system, $R$ is the projected Galactocentric distance, increasing radially outwards, $\phi$ is in the direction of Galactic rotation and $z$ is the same as $z$ in the Cartesian system.
In the Galactocentric spherical system, $r$ is the Galactocentric distance, increasing radially outwards, $\theta$ points toward the South Galactic Pole and $\phi$ is in the direction of Galactic counter-rotation.

\section{Potential Applications and perspectives of the SMSS red giant sample}
The sample of about one million red giants with stellar atmospheric parameters and distances estimated in this work will be very useful for various Galactic studies, including characterizing the structure, chemical and kinematical properties of the MW as well as identifying tidal streams and debris of disrupted dwarf galaxies and star clusters. 
Those studies are important for advancing our understanding of the formation and evolution of our Galaxy.
The sample will be accessible at \url{https://yanghuang0.wixsite.com/yangh/research}.

To show the power of this sample, we present the median metallicity map of our Galaxy in the $R$--$z$ plane (see Fig.\,18). 
This metallicity map covers the currently largest extent of our Galaxy and is 3-4 times bigger than that from SDSS (Ivezi{\'c} et al. 2008).
In the map, we can easily recognize the thin, thick and halo populations, although significant population effects have not been corrected for.\footnote{As Fig.\,5 shows, red giant stars of different colors and metallicities have different absolute magnitudes. Thus, for a given limiting magnitude of SMSS DR1.1, samples of red giant stars of different populations probe different volumes depths -- that of the blue metal-rich population probes shallower than that of the red metal-poor population.
These population effects have not been corrected for the metallicity map presented in Fig.\,18.}
A more detailed analysis of this map will be presented in a separate paper.

At present, our sample does not cover the Galactic plane since most of the stars in SMSS DR1.1 are without $uv$ magnitudes.
In addition, the current sample is mostly limited to 20-30\,kpc given the shallow magnitude limits of SMSS DR1.1.
However, these two shortcomings of the current sample will be overcome by the forthcoming SMSS DR2.
The photometry of DR2 is expected to be improved significantly compared to the DR1 for fields nearer to the Galactic plane, especially in the $uv$ bands.
In addition, the limiting magnitude of DR2 will be about 2\,mag fainter than that of the DR1 over more than one third of the hemisphere.
In the long run, we are also looking forward to an even more fascinating, ambitious project in the northern sky -- the Multi-channel Photometric Survey Telescope (Mephisto\footnote{\url{http://www.swifar.ynu.edu.cn/info/1015/1073.htm}}).
Mephisto is designed to have a 1.6 m primary mirror and a 3.14 deg$^2$ field of view, equipped with three CCD cameras, with a total of 1.4 Giga pixels.
Mephisto will adopt the SkyMapper filter set, and hence provide the opportunity to construct a truly all-sky view of the Milky Way, when combined with SkyMapper in the South. Hence, in the near future, we will have precise measurements of stellar atmospheric parameters and distances for several billion stars down to $r \sim$\,21--22 mag across the whole sky.

Before summarizing, it is worth mentioning the independent work by C19 again.
It has some similarities with the current work, e.g., deriving stellar effective temperatures and metallicities from the SMSS DR1.1 photometry.
However, we note that their photometric calibration only can work for [Fe/H] down to $-2.0$, whereas ours can work for [Fe/H] down to $-2.6$.
Compared to the APOGEE metallicities, our photometric metallicities (with a scatter of 0.17\,dex) are more preciser than those of C19 (with a scatter of 0.25\,dex).
In addition, we have derived distances for stars without good Gaia parallax measurements while C19 do not give distance estimates for distant stars.

\section{Summary}
Using training data sets from the common stars between SMSS DR1.1 and the LAMOST Galactic surveys, we have developed photometric methods to select red giants and determine their stellar atmospheric parameters, i.e. effective temperature $T_{\rm eff}$, surface gravity log\,$g$ and metallicity [Fe/H].
The distances of the stars are estimated with two different approaches.
For nearby bright stars ($d \leq$\,4.5\,kpc), we adopt distance estimates from Gaia DR2 parallax measurements.
For distant faint stars ($d > 4.5$\,kpc), we estimate distances by deriving their absolute magnitudes from $(g-i)_0$ colors and photometric metallicities.

From various tests, we estimate that the stellar atmospheric parameters estimated by our photometric methods are better than $\sim80$\,K for $T_{\rm eff}$,  $\sim0.18$\,dex for [Fe/H] and $\sim0.35$\,dex for log\,$g$.
For the surface gravity log\,$g$, we also note a significant systematic error (about 0.25\,dex) for log\,$g$ around 2.3 due to the poor sensitivities of SMSS colors on separation RCs and RGBs.
For the distances estimated from $(g-i)_0$ colors and photometric metallicities, test with member stars of Globular clusters show a median uncertainty of 16 per cent with a negligible zero-point offset.

With the developed photometric methods, we have successfully selected nearly one million red giants from the SMSS DR1.1 with good-quality photometry (uncertainties in $uvgi$ smaller than 0.05\,mag) and derive their atmospheric parameters and distances.
The sample stars are mostly within 10\,kpc from the Sun but have a tail reaching out as far as 40\,kpc.
Proper motion measurements from Gaia DR2 and radial velocity measurements from several cross-matched spectroscopic surveys are also collected, and are available for almost all and 44 per cent of the sample, respectively.
This sample will be a very useful sample for various Galactic studies, including characterizing the structure, chemical and kinematical properties of the MW as well as identifying tidal streams and debris of disrupted dwarf galaxies and star clusters.

In the near future, we expect to have precise measurements of stellar atmospheric parameters and distances of several billion red giants down to $r \sim$\,21-22 mag for the whole sky, by the significant improvement in both the limiting magnitudes of all bands and the sky coverage in the $uv$ bands of the SMSS DR2$+$, and the new northern-sky survey project Mephisto.

 \section*{Acknowledgements} 
We would like to thank the referee for his/her helpful comments.
 
The Guoshoujing Telescope (the Large Sky Area Multi-Object Fiber Spectroscopic Telescope, LAMOST) is a National Major Scientific Project built by the Chinese Academy of Sciences. Funding for the project has been provided by the National Development and Reform Commission. LAMOST is operated and managed by the National Astronomical Observatories, Chinese Academy of Sciences.

This work has made use of data from the European Space Agency (ESA) mission Gaia (https://www.cosmos.esa.int/gaia), processed by the Gaia Data Processing and Analysis Consortium (DPAC, https://www.cosmos.esa.int/web/gaia/dpac/consortium).

The national facility capability for SkyMapper has been funded through ARC LIEF grant LE130100104 from the Australian Research Council, awarded to the University of Sydney, the Australian National University, Swinburne University of Technology, the University of Queensland, the University of Western Australia, the University of Melbourne, Curtin University of Technology, Monash University and the Australian Astronomical Observatory. SkyMapper is owned and operated by The Australian National University's Research School of Astronomy and Astrophysics. The survey data were processed and provided by the SkyMapper Team at ANU. The SkyMapper node of the All-Sky Virtual Observatory (ASVO) is hosted at the National Computational Infrastructure (NCI). Development and support the SkyMapper node of the ASVO has been funded in part by Astronomy Australia Limited (AAL) and the Australian  Government through the Commonwealth's Education Investment Fund (EIF) and National Collaborative Research Infrastructure Strategy (NCRIS), particularly the National eResearch Collaboration Tools and Resources (NeCTAR) and the Australian National Data Service Projects (ANDS).

 This work is supported by the National Natural Science Foundation of China U1531244, 11833006, 11811530289, U1731108, 11803029 and 11473001 and the Yunnan University grant No.\,C176220100006 and C176220100007.

\appendix
\section{Empirical extinction coefficients for the SkyMapper passbands}

\begin{table}
\centering
\caption{$R (a-b)$ for various colors.}
\begin{threeparttable}
\begin{tabular}{cccc}
\hline
Color & This work & W18\tnote{c} & Fitzpatrick\tnote{d}  \\
\hline
\multirow{2}{*}{$u-v$}&$0.342 \pm 0.018$\tnote{a}&\multirow{2}{*}{$0.268$}&\multirow{2}{*}{$0.353$}\\
         &$0.197 \pm 0.009$\tnote{b}&&\\
$v-g$&$1.326 \pm 0.024$&$1.040$&$1.174$\\
$g-r$&$0.722 \pm 0.013$&$0.698$&$0.761$\\
$V-r$&$0.413 \pm 0.016$&$0.812$&$0.470$\\
$r-i$&$0.655 \pm 0.017$&$0.700$&$0.812$\\
$i-z$&$0.412 \pm 0.009$&$0.382$&$0.436$\\
\hline
\end{tabular}
\begin{tablenotes}
\item[a] $R (u-v)$ derived by stars with $0.20 \le g-i < 0.8$. 
\item[b] $R (u-v)$ derived by stars with $0.80 \le g-i \le 1.5$.
\item[c] Predictions by an $R_{V} = 3.1$ Fitzpatrick extinction law for flat spectra, adopted from W18.
\item[d] Predictions by an $R_{V} = 3.1$ Fitzpatrick extinction law at $E(B-V) = 0.4$ for a 5750\,K source spectrum (with log\,$g$\,=\,4.5 and solar metallicity), calculated in this work.
\end{tablenotes}
\end{threeparttable}
\end{table}

\begin{table}
\centering
\caption{$R (a)$ for SkyMapper passbands.}
\begin{threeparttable}
\begin{tabular}{cccc}
\hline
Passband & This work\tnote{a} & W18\tnote{d} & Fitzpatrick\tnote{d} \\
\hline
\multirow{2}{*}{$u$}&$5.075 \pm 0.018$\tnote{b}&\multirow{2}{*}{$4.294$}&\multirow{2}{*}{$4.932$}\\
         &$4.930 \pm 0.009$\tnote{c}&&\\
$v$&$4.733 \pm 0.024$&$4.026$&$4.579$\\
$g$&$3.407 \pm 0.013$&$2.986$&$3.404$\\
$r$&$2.685 \pm 0.016$&$2.288$&$2.643$\\
$i$&$2.030 \pm 0.017$&$1.588$&$1.831$\\
$z$&$1.618 \pm 0.009$&$1.206$&$1.396$\\
\hline
\end{tabular}
\begin{tablenotes}
\item[a] Calculated using $R(V) = 3.1$ and reddening coefficients from the 2nd column of Table\,A1.
\item[b] $R (u)$ derived by stars with $0.20 \le g-i < 0.8$. 
\item[c] $R (u)$ derived by stars with $0.80 \le g-i \le 1.5$.
\item[d]  Same as Table\,A1 but for $R (a)$.
\end{tablenotes}
\end{threeparttable}
\end{table}

\begin{figure*}
\begin{center}
\includegraphics[scale=0.50,angle=0]{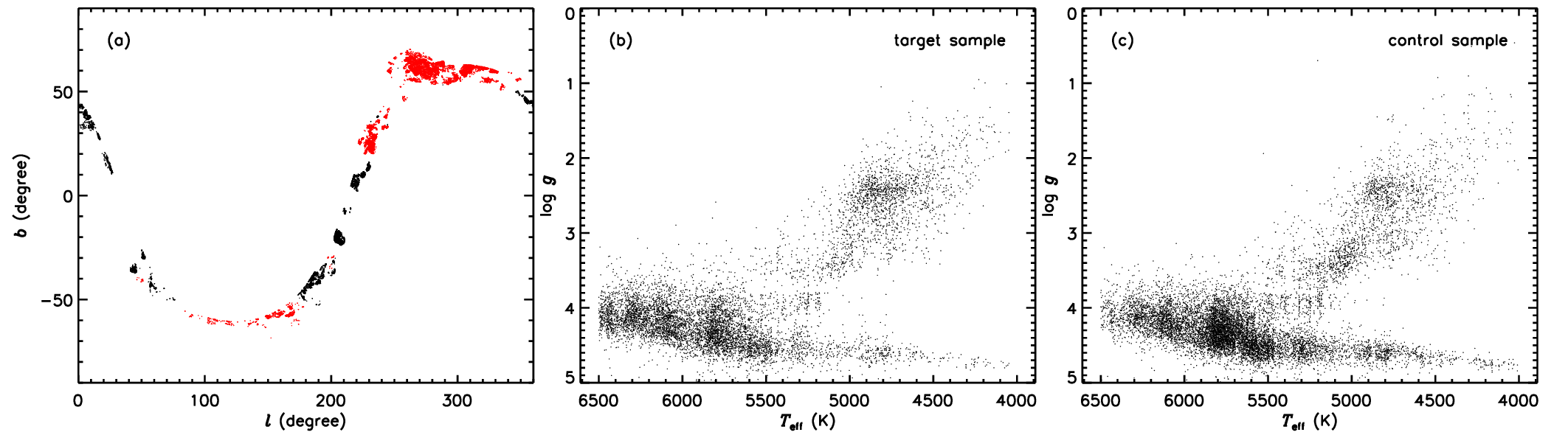}
\caption{{\it Left panel:} Spatial coverage of the control (red dots) and target sample (black dots) samples in Galactic coordinates; {\it Middle panel:} $T_{\rm eff}$--log\,$g$ diagram of the target sample; {\it Right panel:} $T_{\rm eff}$--log\,$g$ diagram of the control sample.}
\end{center}
\end{figure*}

\begin{figure*}
\begin{center}
\includegraphics[scale=0.45,angle=0]{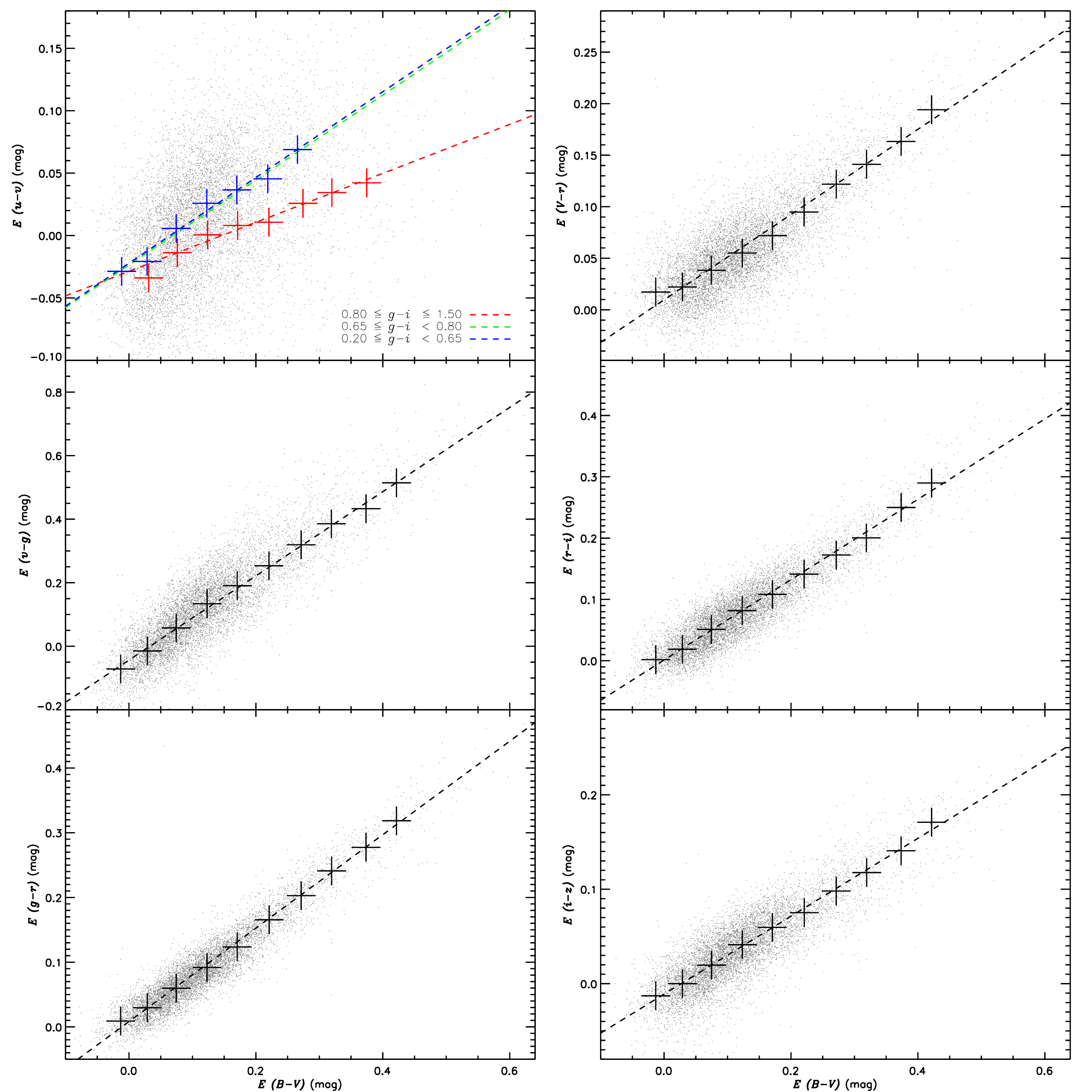}
\caption{Reddening coefficients for the colors $u-v$, $v-g$, $g-r$, $V-r$, $r-i$ and $i-z$ deduced from the target sample for the SkyMapper and APASS passbands. 
Black dots denote data deduced from the individual stars.
Except for the color $u-v$, large black plus signs denote median values obtained by binning the data points into ten groups with a bin size of 0.05\,mag in $E(B-V)$. 
The median values in the individual bins are calculated with a 3$\sigma$ clipping procedure.
The black dashed lines are first-order polynomial fits to the black plus signs, where each point carries equal weight. 
For the color $u-v$, the calculations are the same as for other colors except that stars are grouped into different color bins: $0.20 \leq g-i < 0.65$ (blue symbols), $0.65 \leq g-i < 0.80$ (green symbols) and $0.80 \leq g-i \leq 1.50$ (red symbols).
The reddening coefficients of the color $u-v$ deduced from stars of color bins $0.20 \leq g-i < 0.65$ and $0.65 \leq (g-i) < 0.8$ are almost the same and we have therefore adopted their mean value for both of the two color bins (see Tables\,A1 and A2).}
\end{center}
\end{figure*}

In this Section, we derive the empirical extinction coefficients for the SkyMapper passbands using the ``star pair" technique developed by Yuan, Liu \& Xiang (2013, hereafter YLX13).
Doing so, we first define the control sample and target sample, using common stars between SMSS DR1.1 and the LAMOST Galactic surveys.
For the control sample, we require that the stars satisfy the following criteria:
\begin{itemize}[leftmargin=*]

\item LAMOST spectral SNR greater than 30;

\item $4000 \leq T_{\rm eff} < 6500$\,K, $0.0 < $log\,$g$\,$< 5.0$ and $-1.5 <$\,[Fe/H]\,$< 0.5$;

\item SkyMapper photometric uncertainties in all the six $uvgriz$ bands smaller than 0.035\,mag;

\item $BV$ photometry available from the APASS DR9 and uncertainties smaller than 0.05\,mag;

\item Galactic latitudes $|b| \ge 20^{\circ}$ and SFD98 $E(B-V) \leq 0.03$\,mag.

\end{itemize}
For the target sample, the first four criteria are same as for the control sample but the last criterion changes SFD98 $E(B-V) \geq 0.10$\,mag.
With above criteria, a total of 11518 and 8210 stars are selected for the control and target sample, respectively.
The spatial coverage and distribution in the $T_{\rm eff}$--log\,$g$ plane of the two samples are shown in Fig.\,A1.

For each target star, we select its control stars from the control sample by requiring that their values of $T_{\rm efff}$, log\,$g$ and [Fe/H] that differ from those of the target star within 150\,K, 0.25\,dex and 0.10\,dex, respectively.
The intrinsic colors of the target star are then derived assuming that the intrinsic colors (of both target and control stars) vary linearly with $T_{\rm eff}$, log\,$g$ and [Fe/H], a reasonable assumption considering the small ranges of parameters involved.
For a given color, the reddening value of the target star is then measured as the difference between the observed and intrinsic colors.
The control sample is corrected for reddening using SFD98 $E(B-V)$ values (assuming overestimated by 15.5 per cent, see Section\,2) and an initial set of reddening coefficients from W18 (for the $uvgriz$ bands, see Table\,A1) and Fitzpatrick (1999; for the $BV$ bands).
A new set of reddening coefficients is then derived by comparing the color excesses relative to $E(B-V)$ for the target samples.
We iterate the whole process until the derived set of reddening coefficients is in agreement with the one used to deredden the control sample.

The final results of the reddening coefficients for the colors $u-v$, $v-g$, $g-r$, $V-r$, $r-i$, $i-z$, as well as for $V - r$ for the target sample of SkyMapper are presented in Table\,A1 and shown in Fig.\,A2.
The details of the calculations are the same as in YLX13.
Here, we note that the reddening coefficient of the color $u-v$ is quite sensitive to the color itself, in particular for stars of $g-i \ge 0.80$. 
This is, at least partly, related to the red leak of the $u$ band (Bessell et al. 2011; W18).
Assuming $R_{V} = \frac{A_{V}}{E(B-V)} = 3.1$, the extinction coefficients for all the six SkyMapper passbands are also derived and are presented in Table\,A2.

\end{document}